\documentclass[12pt, preprint]{aastex}

\shorttitle{M37 Transit Survey}
\shortauthors{Hartman et al.}

\begin{document}

\title{Deep MMT\footnote{Observations reported here were obtained at the MMT Observatory, a joint facility of the Smithsonian Institution and the University of Arizona.} Transit Survey of the Open Cluster M37 IV: Limit on the Fraction of Stars With Planets as Small as $0.3 R_{J}$}
\author{J.~D.~Hartman\altaffilmark{2}, B.~S.~Gaudi\altaffilmark{3}, M.~J.~Holman\altaffilmark{2}, B.~A.~McLeod\altaffilmark{2}, K.~Z.~Stanek\altaffilmark{3}, J.~A.~Barranco\altaffilmark{4}, M.~H.~Pinsonneault\altaffilmark{3}, S.~Meibom\altaffilmark{2}, and J.~S.~Kalirai\altaffilmark{5,6}}
\altaffiltext{2}{Harvard-Smithsonian Center for Astrophysics, 60 Garden St., Cambridge, MA~02138, USA; jhartman@cfa.harvard.edu, mholman@cfa.harvard.edu, bmcleod@cfa.harvard.edu, smeibom@cfa.harvard.edu}
\altaffiltext{3}{Department of Astronomy, The Ohio State University, Columbus, OH~43210, USA; gaudi@astronomy.ohio-state.edu, pinsono@astronomy.ohio-state.edu, kstanek@astronomy.ohio-state.edu}
\altaffiltext{4}{Department of Physics and Astronomy, San Francisco State University, 1600 Holloway Ave., San Francisco, CA~94132, USA; barranco@stars.sfsu.edu}
\altaffiltext{5}{University of California Observatories/Lick Observatory, University of California at Santa Cruz, Santa Cruz CA, 95060; jkalirai@ucolick.org}
\altaffiltext{6}{Hubble Fellow}

\begin{abstract}
We present the results of a deep ($15 \la r \la 23$), 20 night survey for transiting planets in the intermediate age open cluster M37 (NGC 2099) using the Megacam wide-field mosaic CCD camera on the 6.5m MMT. We do not detect any transiting planets among the $\sim 1450$ observed cluster members. We do, however, identify a $\sim 1~R_{J}$ candidate planet transiting a $\sim 0.8~M_{\odot}$ Galactic field star with a period of $0.77~{\rm days}$. The source is faint ($V = 19.85~{\rm mag}$) and has an expected velocity semi-amplitude of $K \sim 220~{\rm m/s}~(M/M_{J})$. We conduct Monte Carlo transit injection and recovery simulations to calculate the $95\%$ confidence upper limit on the fraction of cluster members and field stars with planets as a function of planetary radius and orbital period. Assuming a uniform logarithmic distribution in orbital period, we find that $< 1.1\%$, $< 2.7\%$ and $< 8.3\%$ of cluster members have $1.0~R_{J}$ planets within Extremely Hot Jupiter (EHJ, $0.4 < P < 1.0~{\rm day}$), Very Hot Jupiter (VHJ, $1.0 < P < 3.0~{\rm day}$) and Hot Jupiter (HJ, $3.0 < P < 5.0~{\rm day}$) period ranges respectively. For $0.5~R_{J}$ planets the limits are $< 3.2\%$, and $< 21\%$ for EHJ and VHJ period ranges, while for $0.35~R_{J}$ planets we can only place an upper limit of $< 25\%$ on the EHJ period range. For a sample of $7814$ Galactic field stars, consisting primarily of FGKM dwarfs, we place $95\%$ upper limits of $< 0.3\%$, $< 0.8\%$ and $< 2.7\%$ on the fraction of stars with $1.0~R_{J}$ EHJ, VHJ and HJ assuming the candidate planet is not genuine. If the candidate is genuine, the frequency of $\sim 1.0 R_{J}$ planets in the EHJ period range is $0.002\% < f_{EHJ} < 0.5\%$ with $95\%$ confidence. We place limits of $< 1.4\%$, $< 8.8\%$ and $< 47\%$ for $0.5~R_{J}$ planets, and a limit of $< 16\%$ on $0.3~R_{J}$ planets in the EHJ period range. This is the first transit survey to place limits on the fraction of stars with planets as small as Neptune.
\end{abstract}


\section{Introduction}\label{sec:intro}

The discovery by \citet{Mayor.95} of a planet with half the mass of Jupiter orbiting the solar-like star 51 Pegasi with a period of only $4.23~{\rm days}$ shocked the astronomical community. The existence of such a ``hot Jupiter'' (HJ) defied the prevailing theories of planet formation which had been tailored to explain the architecture of the Solar System. Since then, radial velocity (RV) surveys for planets orbiting nearby F, G and K main-sequence stars have determined that $1.2\% \pm 0.2\%$ of these stars host a HJ \citep[][where a HJ is defined as a planet roughly the size of Jupiter that orbits within $0.1~{\rm AU}$ of its star]{Marcy.05}, with indications that this frequency depends on the metallicity of the host star \citep[such that the frequency is roughly proportional to $10^{2{[}Fe/H{]}}$,][]{Fischer.05}. 

Over the last decade there have been numerous surveys for extra-solar planets following a variety of techniques \citep[e.g.][]{Butler.06} with the goal of determining the planet occurrence rate in new regions of parameter space. A successful technique has been to conduct photometric searches for planets that transit their host stars. This technique is particularly sensitive to planets on close-in orbits. To date more than 50 planets have been discovered by this technique\footnote{http://exoplanet.eu}, including numerous very hot Jupiters (VHJ) with orbital periods between 1 and 3 days. \citet{Gaudi.05} used the four transiting planets discovered at the time by the OGLE collaboration \citep{Udalski.02a, Konacki.03, Bouchy.04, Konacki.04, Pont.04} to determine that only $0.1 - 0.2\%$ of FGK stars host a VHJ. \citet{Gould.06} conducted a thorough analysis of the OGLE survey to determine that the frequency of VHJs is $f_{VHJ} = (1/710)(1^{+1.10}_{-0.54})$ and $f_{HJ} = (1/320)(1^{+1.37}_{-0.59})$, while \citet{Fressin.07} found $f_{VHJ} = (1/560)$ and $f_{HJ} = (1/320)$. The SWEEPS survey for transiting planets in the Galactic bulge conducted with the Hubble Space Telescope \citep{Sahu.06} identified a putative class of ultra-short-period planets, or Extremely Hot Jupiters (EHJ) with periods less than $1.0~{\rm day}$ orbiting stars lighter than $0.88 M_{\odot}$. They conclude that $\sim 0.4\%$ of bulge stars more massive than $\sim 0.44 M_{\odot}$ are orbited by a Jupiter-sized planet with a period less than $4.2~{\rm days}$, though they estimate that this fraction is uncertain by a factor of 2. Note that due to their faintness the majority of the SWEEPS candidates are unconfirmed with RV follow-up. In addition to these two surveys, the TrES \citep[e.g][]{Alonso.04}, HAT \citep[e.g.][]{Bakos.07}, XO \citep[e.g.][]{McCullough.06}, and WASP \citep[e.g.][]{CollierCameron.06} surveys have all discovered planets orbiting relatively bright stars in the Galactic field, though to date these surveys have not been used to calculate the planet occurrence frequency.

While photometric surveys of Galactic field stars have been quite successful at finding transiting planets over the last few years, it is generally difficult to measure the planet occurrence frequency with these surveys \citep[for discussions of how this can be done despite the difficulties see][]{Gould.06,Fressin.07,Gaudi.07,Beatty.08}. The difficulty arises from the uncertainty in the parameters (mass, radius, metallicity) of the surveyed stars. Moreover, typical field surveys yield numerous false positives that are often culled in part by eye, these culling procedures are generally difficult to model in determining the detection efficiency of the survey. In contrast to field surveys, surveys of globular and open star clusters observe a population of stars with parameters that are relatively easy to determine en masse, moreover many of the false positive scenarios are less common for this type of survey. There has been significant work invested in developing optimum strategies to search for planets in stellar clusters \citep{Janes.96,vonBraun.05,Pepper.05}. A number of groups have completed transit surveys of open clusters, including the UStAPS \citep{Street.03,Bramich.05,Hood.05}, EXPLORE-OC \citep{vonBraun.05}, PISCES \citep{Mochejska.05, Mochejska.06}, STEPSS \citep[][hereafter B06]{Burke.06} and MONITOR \citep{Aigrain.07} projects and a survey by \citet{Montalto.07}. There have also been several surveys of globular clusters \citep{Gilliland.00, Weldrake.05, Weldrake.08}.

While to date no confirmed transiting planet has been found in a stellar cluster, many of these surveys have placed limits on the frequency of hot transiting planets, typically as functions of planetary radius as well as period. The globular cluster surveys have placed the most stringent constraints; the null result for the core of 47~Tucanae by \citet{Gilliland.00} suggests that the frequency of HJ in this environment is at least an order of magnitude less than for the solar neighborhood, while the null result for the outskirts of the same cluster by \citet{Weldrake.05} is inconsistent with the planet frequency in the solar neighborhood at the $3.3\sigma$ level and suggests that the dearth of planets in this globular cluster is due to low metallicity rather than crowding effects. The open cluster surveys, on the other hand, have typically placed limits on the occurrence frequency that are well above the $1.2\%$ measured by the RV surveys. Notably B06 conducted a thorough Monte Carlo simulation of their transit survey of the open cluster NGC 1245 to limit the frequency of EHJ, VHJ and HJ with radii of $1.5~R_{J}$ to $< 1.5\%$, $< 6.4\%$ and $< 52\%$ respectively. The fundamental limit on the ability of open cluster surveys to place meaningful limits on the occurrence frequency of Jupiter-sized planets appears to be due to the relatively small number of stars in an open cluster. B06 find that for their survey strategy, $\sim 7400$ dwarf stars would have to be observed for at least a month to put a limit of less than $2\%$ on the planet frequency, which is significantly larger than the typical size of an open cluster. One open cluster that has been a popular target is NGC 6791. This cluster is old \citep[$t \sim 8~{\rm Gyr}$,][]{Carraro.06, Kalirai.07}, metal rich \citep[${[}M/H{]} \sim +0.4$,][]{Gratton.06, Origlia.06} and contains a large number of stars \citep[$M > 4000 M_{\odot}$,][]{Kaluzny.92}, though it is also very distant \citep[$(m-M)_{0} \sim 12.8$,][]{Stetson.03} so that lower main sequence stars in the cluster are quite faint. It has been the target of three transit searches \citep{Bruntt.03,Mochejska.05,Montalto.07}, the most recent of which found that their null result is inconsistent at the $\sim 95\%$ level with the RV HJ frequency at high metallicity.

The paucity of stars in open clusters appears to limit their usefulness as probes of the HJ frequency \citep[excluding, perhaps, the result from][]{Montalto.07}. They may, however, be useful for probing smaller planet radii \citep[see][]{Pepper.06a}. In the last several years RV surveys have discovered a number of Neptune and super-Earth-mass planets \citep[HN, $M < 0.1 M_{J}$;][]{Butler.04,Endl.07,Fischer.07,Lovis.06,Melo.07,Rivera.05,Santos.04a,Udry.05,Udry.07,Vogt.05}. One of these planets, GJ 436 b, has been discovered to transit its host star \citep{Gillon.07}. Little, however, is known about the frequency of these planets. Determining, or placing meaningful limits on this frequency would provide a powerful test of planet formation models. The theoretical predictions of the frequency of these objects run the gamut from a steep decline in the frequency of HN relative to HJ \citep{Ida.04}, except perhaps for M-dwarfs \citep{Ida.05}, to HN being ubiquitous \citep{Brunini.05}. 

In this paper, the fourth and final in a series, we present the results of a survey for transiting hot planets as small as Neptune in the intermediate age open cluster M37 (NGC 2099) using the MMT. We were motivated to conduct this transit survey by \citet{Pepper.05,Pepper.06a} who suggested that it may be possible to find Neptune-sized planets transiting solar-like stars by surveying an open cluster with a large telescope. The Megacam mosaic imager on the MMT \citep{McLeod.00} is an ideal facility for conducting such a survey due to its wide field of view and deep pixel wells that oversample the stellar point spread function (PSF). Preliminary observations of NGC 6791 suggested that finding Neptune-sized planets was indeed technically feasible using this facility \citep{Hartman.05}. Using the formalism developed by \citet{Pepper.05} we found that M37 is the optimum target for MMT/Megacam to maximize the number of stars to which we would be sensitive to Neptune-sized planets. We note that a drawback of this type of survey is that any identified candidates will be quite faint making follow-up RV confirmation difficult. For planets significantly smaller than $1.0 R_{J}$, false positives where the transiting object is a small star or brown dwarf are no longer applicable. Given the depth of the survey very few giants will be included in the sample, and those that are, can easily be rejected based on their colors. Low-precision spectroscopy may be sufficient to rule out various blended binary scenarios. Therefore, it is reasonable to suppose that for a given small-radius candidate one could make a strong argument that the object is a real planet without obtaining a RV determination of its mass. Also note that similar difficulties will be faced by the \emph{Corot} and \emph{Kepler} space missions (albeit for even smaller planets), so our experiment may provide a useful analogy to these missions.

We conducted the survey over twenty nights between December, 2005 and January, 2006, accumulating more than 4000 quality images of the cluster. This is easily the largest telescope ever utilized for such a survey. In the first paper in the series \citep[Paper I]{Hartman.08a} we describe the observations and data reduction, combine photometric and spectroscopic data to refine estimates for the cluster fundamental parameters ($t = 550 \pm 30~{\rm Myr}$ with overshooting, $[M/H] = +0.045 \pm 0.044$, $(m-M)_{V} = 11.57 \pm 0.13~{\rm mag}$ and $E(B-V) = 0.227 \pm 0.038~{\rm mag}$), and determine the cluster mass function down to $0.3 M_{\odot}$. In the second paper \citep[Paper II]{Hartman.08b} we analyze the light curves of $\sim 23000$ stars observed by this survey to discover 1430 variable stars. In the third paper \citep[Paper III]{Hartman.08c} we use the light curves to measure the rotation periods of 575 probable cluster members. This is the largest sample of rotation periods for a cluster older than a few hundred Myr, and thus provides a unique window on the late time rotation evolution of low-mass main sequence stars.

In the following section we will summarize our observations and data reduction. In \S~\ref{sec:transselect} we discuss the pipeline used to remove systematic variations from light curves and identify candidate transiting planets. In \S~\ref{sec:transcand} we describe the candidate transiting planets identified by this survey, finding no candidates that are probable cluster members. In \S~\ref{sec:deteff} we conduct Monte Carlo simulations to determine the transit detection efficiency of our survey. In \S~\ref{sec:results} we present our results on the limit of stars with planets for various planetary radii and orbital periods. Finally, we conclude in \S~\ref{sec:discussion}.

\section{Summary of Observations and Data Reduction}\label{sec:sumobs}

The observations and data reduction procedure were described in detail in Papers I and II, we provide a brief overview here. The photometric observations consist of $gri$ photometry for $\sim 16,000$ stars, $gri$ photometry for stars in a field located two degrees from the primary M37 field and at the same Galactic latitude, and $r$ time-series photometry for $\sim 23,000$ stars, all obtained with the Megacam instrument \citep{McLeod.00} on the 6.5 m MMT. Megacam is a $24\arcmin \times 24\arcmin$ mosaic imager consisting of 36 2k$\times$4k, thinned, backside-illuminated CCDs that are each read out by two amplifiers. The mosaic has an unbinned pixel scale of $0\farcs 08$ which allows for a well sampled point-spread-function (PSF) even under the best seeing conditions. To decrease the read-out time we used $2 \times 2$ binning with the gain set so that the pixel sensitivity became non-linear before the analog-to-digital conversion threshold of 65,536 counts. Because of the fine sampling and the relatively deep pixel wells, one can collect $2\times 10^7$ photons in $1\arcsec$ seeing from a single star prior to saturation, setting the photon limit on the precision in a single exposure to $\sim$ 0.25 mmag.

The primary time-series photometric observations consist of $\sim 4000$ high quality images obtained over twenty four nights (including eight half nights) between December 21, 2005 and January 21, 2006. We obtained light curves for stars with $14.5 \la r \la 23$ using a reduction pipeline based on the image subtraction technique and software due to \citet{Alard.98} and \citet{Alard.00}. The resulting light curves were passed through the processing and transit detection pipeline that we describe in the following section. We used the {\scshape Daophot} 2 and {\scshape Allstar} PSF fitting programs and the {\scshape Daogrow} program \citep{Stetson.87,Stetson.90,Stetson.92} to obtain the $g$, $r$, and $i$ single-epoch photometry.

As described in Paper I we also take $BV$ photometry for stars in the field of this cluster from \citet{Kalirai.01}, $K_{S}$ photometry from 2MASS \citep{Skrutskie.06} and we transform our $ri$ photometry to $I_{C}$ using a transformation based on the $I_{C}$ photometry from \citet{Nilakshi.02}.

In addition to the photometry, we also obtained high-resolution spectroscopy of 127 stars using the Hectochelle multi-fiber, high-dispersion spectrograph \citep{Szentgyorgyi.98} on the MMT. The spectra were obtained on four separate nights between February 23, 2007 and March 12, 2007 and were used to measure $T_{eff}$, $[Fe/H]$, $v\sin i$ and the radial velocity (RV) via cross-correlation against a grid of model stellar spectra computed using ATLAS 9 and SYNTHE \citep{Kurucz.93}. The classification procedure was developed by Meibom et al. (2008, in preparation), and made use of the \emph{xcsao} routine in the {\scshape Iraf}\footnote{{\scshape Iraf} is distributed by the National Optical Astronomy Observatories, which is operated by the Association of Universities for Research in Astronomy, Inc., under agreement with the National Science Foundation.} \emph{rvsao} package \citep{Kurtz.98} to perform cross-correlation. We use these spectra to provide stellar parameters and radial velocity measurements for several of the host stars to the candidate transiting planets.

\section{Transit Selection Pipeline}\label{sec:transselect}

In this section we describe the pipelines used to remove systematic variations from the light curves and select candidate transiting planets. We use these pipelines first to identify candidates (\S~\ref{sec:transcand}) and then to calibrate the detection efficiency by conducting transit injection and recovery simulations (\S~\ref{sec:deteff}). The pipeline includes several steps: post-processing the light curves, devising transit selection criteria, and applying the selection criteria to the data to find candidates. We have devised three distinct pipelines, which we refer to as selection criteria sets 1-3, that differ in the number of post-processing routines applied and the manner in which candidates are selected. We first discuss the post-processing routines and the precision of the resulting post-processed light curves, we then discuss the selection criteria.

\subsection{Light Curve Post-Processing}\label{sec:lcproc}

\begin{figure}[p]
\plotone{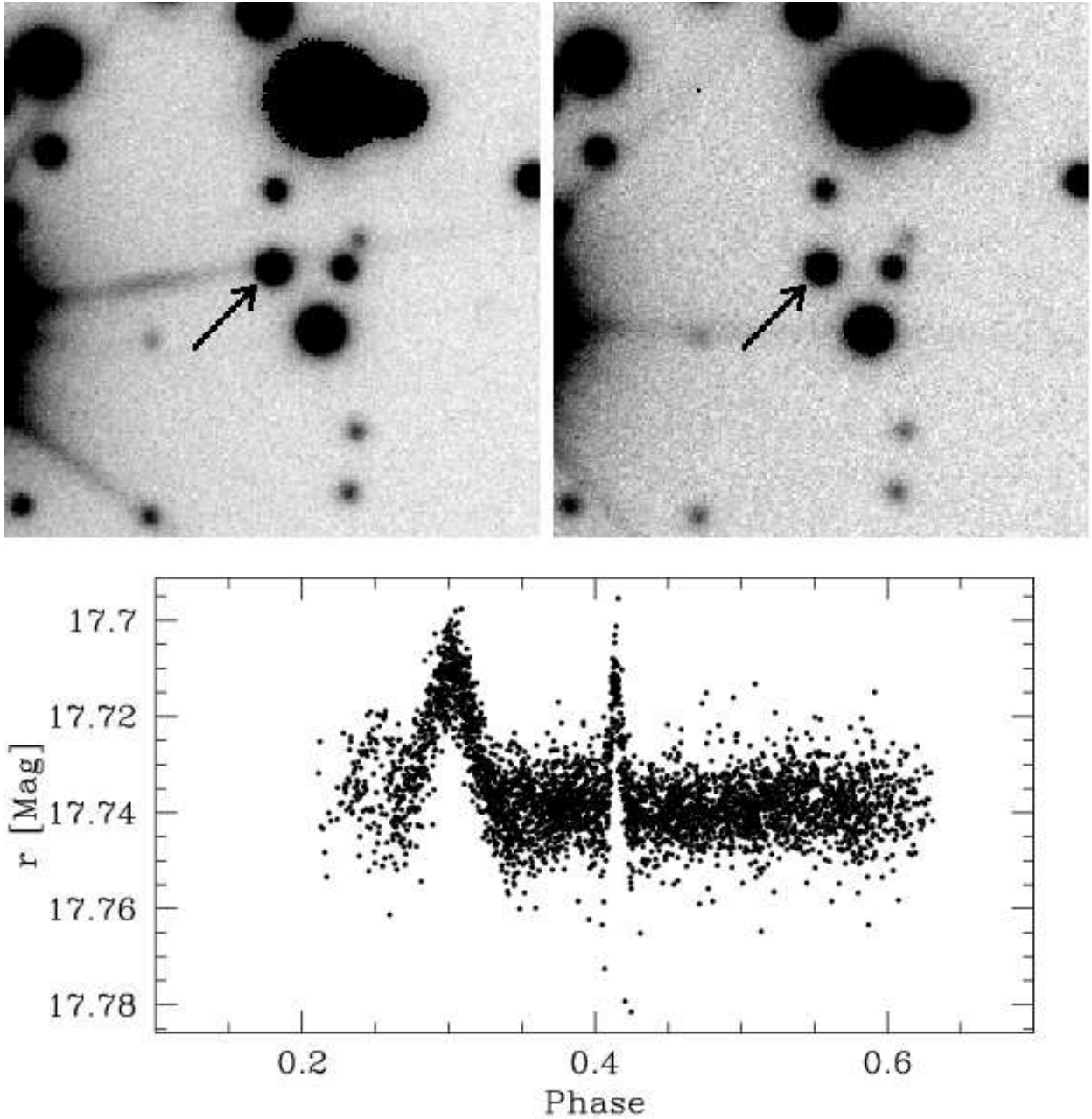}
\caption{Example of a star with a light curve that shows repeated brightenings due to diffraction spikes from a neighboring bright star. The upper two panels show the star in question on a $0\farcm 5 \times 0\farcm 5$ FOV, when a diffraction spike is over the star (left) and when no spike is over the star (right). The two images were taken 100 minutes apart. The bottom panel shows the light curve of the star phased at 1 sidereal day, note the two periodic brightenings caused by the rotation of diffraction spikes in the images over the course of each night.}
\label{fig:diffractionspike}
\end{figure}

\begin{figure}[p]
\epsscale{0.7}
\plotone{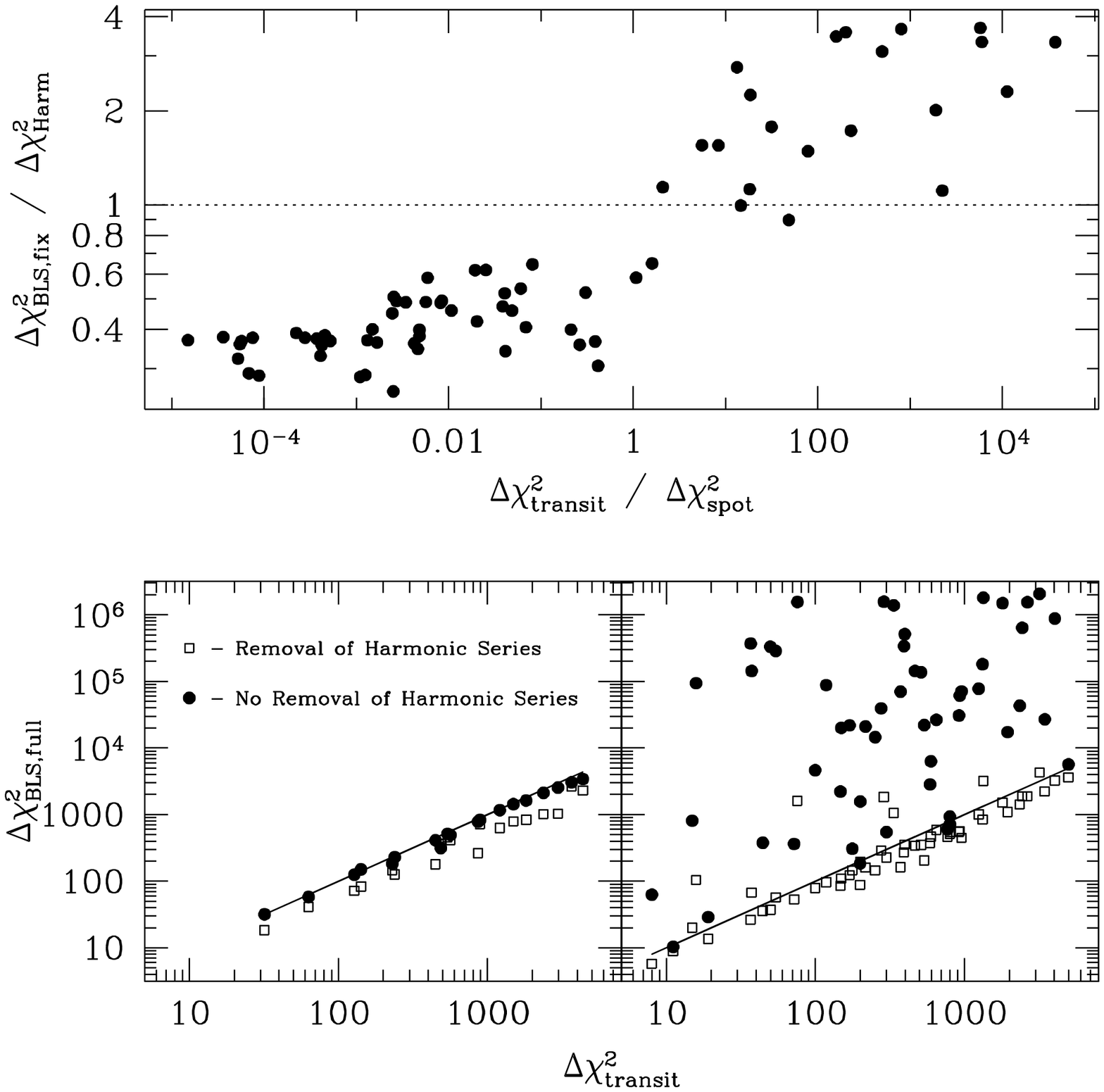}
\caption{Transit recovery results for model light curves including both transits and photometric variations due to starspots. (Top) Comparison between the ratio of the model transit strength to the model spot strength, measured as a contribution to $\chi^{2}$ for each simulated signal, and the ratio of the best boxcar fit to the best harmonic series fit at the peak $L-S$ period. We divide the data into two classes, those with $\Delta \chi^{2}_{{\rm BLS,fix}} / \Delta \chi^{2}_{{\rm Harm}} > 1$ (i.e. a transit model fits the light curve better than a spot model), and those with $\Delta \chi^{2}_{{\rm BLS,fix}} / \Delta \chi^{2}_{{\rm Harm}} < 1$ (i.e. a spot model fits the light curve better than a transit model). (Bottom) Comparison between the best-fit BLS transit model, using a full period search, and the strength of the injected transit. We show the results separately for simulations with $\Delta \chi^{2}_{{\rm BLS,fix}} / \Delta \chi^{2}_{{\rm Harm}} > 1$ (left) and simulations with $\Delta \chi^{2}_{{\rm BLS,fix}} / \Delta \chi^{2}_{{\rm Harm}} < 1$ (right). In each plot we show the results for both removing and not removing a best-fit harmonic series from the light curve before running the full BLS search. The solid lines show the ideal relation where $\Delta \chi^{2}_{{\rm BLS,full}} = \Delta \chi^{2}_{{\rm transit}}$. When the transit model fits the light curve better than a harmonic series (left) the harmonic series should not be removed from the light curve, when the harmonic series fits the light curve better (right) it should be removed before searching for transits.}
\label{fig:simspottransit}
\end{figure}

The raw light curves returned from the image subtraction procedure exhibit substantial scatter due to instrumental artifacts as well as astrophysical variations. Before searching these light curves for low-amplitude transit signals we must take steps to reduce the time-correlated noise. This process may reduce the sensitivity to high S/N planets, so for selection criteria set 3 we only apply a limited post-processing routine. Our post-processing pipeline consists of the following steps:
\begin{enumerate}
\item (selection criteria sets 1-3) We first clip points from each light curve that are more than 5 standard deviations from the mean magnitude. We perform two iterations of this procedure.
\item (selection criteria sets 1-3) For each image $i$, we determine $f_{i}$, the fraction of light curves for which image $i$ is at least a three sigma outlier. Images with $f_{i} > f_{c}$, the cutoff fraction, are removed from all the light curves. This process is performed independently for each of the 36 chips. We choose $f_{c}$ for each chip based on a visual inspection of the histogram of $f_{i}$ values, we use values that range from $3\%$ to $5\%$. This process typically removes $\sim 500$ images from the light curves.
\item (selection criteria sets 1 and 2 only) We then remove 1 sidereal day, or 0.9972696 day period signals from the light curves. This is done to remove artifacts due to, for example, rotating diffraction spikes that have a period of exactly 1 sidereal day. Figure~\ref{fig:diffractionspike} shows an example of a light curve that exhibits brightenings at a period of 1 sidereal day as diffraction spikes from nearby bright stars sweep over the star. The diffraction spikes rotate in the image due to the need to rotate the camera with respect to the secondary mirror supports to keep stars on the same pixels throughout the night on the Alt-Az MMT. The number, width, depth, and shape of the brightenings seen in the light curves vary from star to star, moreover additional artifacts such as color dependent atmospheric extinction will also give rise to apparent variability with a period of 1 sidereal day. We remove these signals by binning the light curves in phase, using 200 bins, and then adjusting the points in a bin by an offset so that the average of the bin is equal to the average of the light curve.
\item (selection criteria sets 1 and 2 only) As discussed in Paper III, we found that $\sim 1/3$ of the probable cluster members that we observed show quasi-periodic variations with amplitudes of $\sim 1\%$ and periods ranging from $0.3$ to $15~{\rm days}$. These variations are due to the presence of large spots on the surfaces of these relatively young ($550~{\rm Myr}$), rapidly rotating stars. While it may be possible to identify a $\sim 1\%$ transit on top of such a signal, it would be increasingly difficult to identify shallower transits without taking steps to remove these variations. Using the Lomb-Scargle algorithm \citep{Lomb.76, Scargle.82, Press.89} we identify the period $P$ of the best-fit sine curve to each light curve. We then fit to the post-step-3 light curve a signal of the form
\begin{equation}
m = A_{0} + A_{1}\sin(2Pt/2\pi + \phi_{1}) + A_{2}\sin(Pt/2\pi + \phi_{2}) + A_{3}\sin(0.5Pt/2\pi + \phi_{3})
\label{eqn:harmseries}
\end{equation}
where the $A_{i}$ and $\phi_{i}$ are free parameters, and we calculate $\Delta \chi^{2}_{{\rm Harm}}$, the reduction in $\chi^{2}$ after subtracting the best-fit model from the light curve. Note that we adopt the convention that a more negative value of $\Delta \chi^{2}$ indicates a better fit of the model to the data. This process increases the sensitivity to shallow transits but decreases the sensitivity to deep transits. We therefore fit a box-car transit signal to the post-step-3 light curve, without subtracting the sine curve, phased at period $P$ using the BLS algorithm \citep{Kovacs.02} and calculate $\Delta \chi^{2}_{{\rm BLS,fix}}$, the reduction in $\chi^{2}$ after subtracting the box-car transit model from the light curve. We subtract equation~\ref{eqn:harmseries} from light curves with $\Delta \chi^{2}_{{\rm Harm}} < \Delta \chi^{2}_{{\rm BLS,fix}}$, i.e. we only apply the correction to light curves for which a harmonic series fits the phased light curve better than a transit signal fits it. To test this technique we simulate light curves including variability due to spots as well as transits. We use the \citet{Dorren.87} spot model and the \citet{Mandel.02} transit model. Figure~\ref{fig:simspottransit} compares the results of a full BLS search on light curves with deep transits relative to spots and vice versa for the two cases of removing and not removing a harmonic signal before running BLS. We see that removing a harmonic series from light curves with $\Delta \chi^{2}_{{\rm Harm}} < \Delta \chi^{2}_{{\rm BLS,fix}}$ yields BLS results that are consistent with the injected transit signal, while not removing the harmonic signal from light curves with $\Delta \chi^{2}_{{\rm Harm}} > \Delta \chi^{2}_{{\rm BLS,fix}}$ yields better results than removing the signal.
\item (selection criteria sets 1 and 2 only) Finally we attempt to remove any remaining instrumental or weather related trends from the post-step-4 light curves using the Trend Filtering Algorithm \citep[TFA][]{Kovacs.05}. This algorithm linearly decorrelates each light curve against a representative sample of other light curves. The trend list for each light curve consists of the other stars on the chip with more than $2500$ points, root-mean-square (RMS) $< 0.1~{\rm mag}$ and that are well outside the photometric radius of the star in question. There are typically $\sim 250$ stars in the trend list for each chip.
\end{enumerate}

\subsubsection{Light Curve Precision}

\begin{figure}[p]
\epsscale{1.0}
\plotone{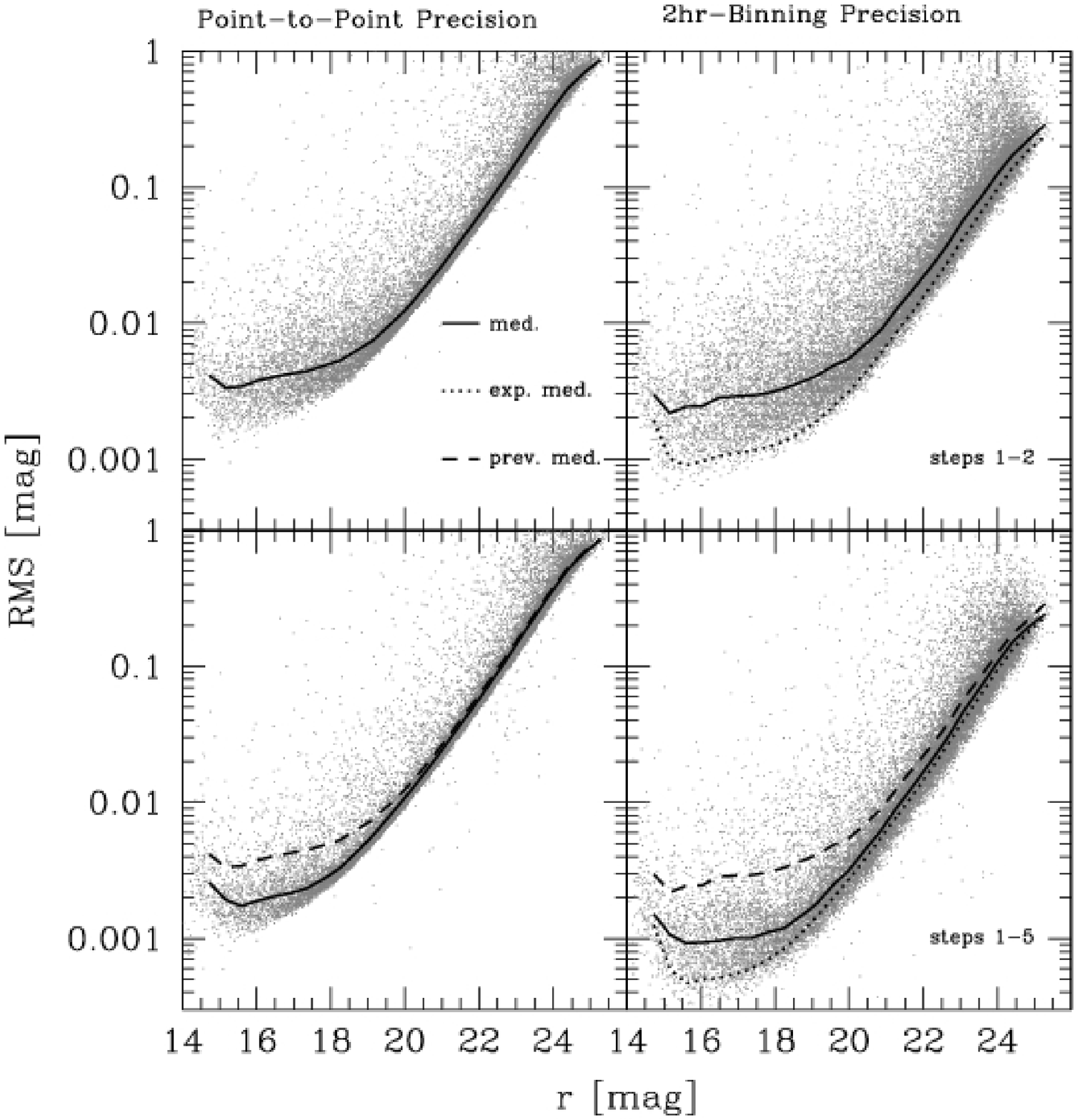}
\caption{Light curve RMS vs. $r$ magnitude for $\sim 22,000$ stars after completing steps 1-2 (top panels) and steps 1-5 (bottom panels) in the light curve processing pipeline. We show both the point-to-point RMS (left) and the RMS after binning the light curves by $2$ hours in time (right), which is approximately the time-scale of a transit by an HJ. The solid lines show the median relation for the data plotted in each panel, while the dashed lines in the bottom two panels show the median relations from the top two panels. The dotted lines show the expected median relation for the binned light curves if the RMS for each light curve were reduced by $N_{b}^{-0.5}$, where $N_{b}$ is the average number of points in a time-bin.}
\label{fig:RMSplot}
\end{figure}

In figure~\ref{fig:RMSplot} we show the RMS as a function of magnitude for the stars after stages 2 and 5 in the light curve processing pipeline. We plot both the point-to-point RMS as well as the RMS after binning the light curves by $2$ hours in time. 

As discussed by \citet{Pont.06}, the limiting factor for detecting low-amplitude transits around bright stars for many transit surveys is the presence of time-correlated systematic variations in the light curves (called red noise, noise that is uncorrelated in time will be referred to as white noise). We can estimate the degree of red noise in our light curves by fitting a noise model to the median RMS-$r$ relation. We use a model of the form:
\begin{equation}
RMS = \sqrt{\left( \frac{2.5}{\ln(10.0)} \right)^{2}\frac{f + s}{f^{2}} + \sigma_{r}^{2}}
\label{eqn:noisemodel}
\end{equation}
where $f = 10^{-0.4(m - z)}$ is the effective number of photo-electrons per image for a source of magnitude $m$ with zero-point $z$, $s = 10^{-0.4(m_{s} - z)}$ is the effective number of photo-electrons per image due to the sky that contaminate a given source and corresponds to a magnitude $m_{s}$, and $\sigma_{r}$ is the effective red noise in magnitudes. Note that $z$ depends on the aperture and efficiencies of the optics and detector, the exposure time, the atmospheric extinction, and the photometric procedure. Also note that $m_{s}$ is the magnitude of a source which has a flux equal to that of the total sky background through the effective photometric aperture. We bin the light curves on time-scales of 5, 10, 30, 60, 120, 180, 240, 1440, 2880, and 7200 minutes and calculate the median RMS-$r$ relation for each binning. We then fit equation~\ref{eqn:noisemodel} simultaneously to all ten relations using the free parameters $z_{T}$, $m_{s}$ and $\sigma_{r,T}$ where $z_{T}$ and $\sigma_{r,T}$ depend on the time-scale, while $m_{s}$ is independent of the time-scale. Table~\ref{tab:RMSmodel} gives the parameters from the model fit to the median RMS-$r$ relations. We also list in table~\ref{tab:RMSmodel} the $r$-magnitude of a source for which the red noise on a given time-scale is equal to the Poisson noise, i.e.
\begin{equation}
\frac{2.5}{\ln(10)}\frac{\sqrt{f_{T} + s_{T}}}{f_{T}} = \sigma_{r,T}.
\end{equation}
As expected, $\sigma_{r,T}$ decreases with increasing $T$. At the time-scales relavent for a transit (1-3 hours), the red noise is $\sim 0.9~{\rm mmag}$. We also note that the values of $z_{T}$ are consistent with the expected values given the gain of the detector and the measured zero-point for the reference image, assuming the effective exposure time for the ``average'' image in the light curve is $\sim 1.0~{\rm minute}$.

%\begin{figure}[p]
%\plotone{f4.eps}
%\caption{The median RMS-$r$ relation for light curves binned at several time-scales (solid lines) are compared with a model relation (equation~\ref{eqn:noisemodel}, dotted lines). From top to bottom, the displayed relations are for binning time-scales of 5, 10, 30, 60, 120, 180, 240, 1440, 2880, and 7200 minutes.}
%\label{fig:RMSmodel}
%\end{figure}

%\begin{figure}[p]
%\epsscale{1.0}
%\plotone{RMSplot_small_errors.eps}
%\caption{Light curve RMS vs. $r$ magnitude for $\sim 22,000$ stars after completing steps 1-5 in the light curve processing pipeline. We show both the point-to-point RMS (left) and the RMS after binning the light curves by $2$ hours in time (right), which is approximately the time-scale of a transit by an HJ. The solid lines show the median relation for the data plotted in each panel. The long-dashed lines show fits of equation~\ref{eqn:noisemodel} to the solid lines holding $z$ constant, while the dot-dashed lines show fits where $z$ is allowed to vary. The dotted lines show the source noise, sky noise and red noise contributions to the dot-dashed relations.}
%\label{fig:RMSplot_small_errors}
%\end{figure}

An alternative method to determine the degree of red noise in the data is to examine the auto-correlation function of the light curves. We find that stars fainter than $r \sim 18.0$ are effectively uncorrelated in time while brighter stars appear to be uncorrelated on time-scales longer than $\sim 200~{\rm minutes}$.

%\begin{figure}[p]
%\epsscale{1.0}
%\plotone{f5.eps}
%\caption{The median autocorrelation function (equation~\ref{eqn:autocorrelation}) of the post-TFA light curves is shown for several magnitude bins. Stars fainter than $r \sim 18.0$ are effectively uncorrelated in time, brighter stars are uncorrelated on time-scales longer than $\sim 200~{\rm minutes}$. Note that the formal uncertainties may be over-estimated by a factor of $\sim 2.0$, so the value of $\rho(T)$ shown here may be too small by a factor of $\sim 4.0$.}
%\label{fig:autocorrelation}
%\end{figure}

\subsection{Transit Selection}\label{sec:transselectsubsection}

To identify the best-fit transit signal in each light curve we use the BLS algorithm. We apply two slight modifications to the algorithm:
\begin{enumerate}
\item Rather than searching over a fixed range of fractional transit width $q = \tau/P$ values (where $\tau$ is the duration of the transit and $P$ is the orbital period), we allow the $q$ range to vary with the trial frequency. We do this by assuming a stellar radius range of $R_{min}$ to $R_{max}$ and taking $q_{min,max} = 0.076\left (R_{min,max} f \right )^{2/3}$ where $f$ is the trial frequency in ${\rm days}^{-1}$, and assuming that $M/M_{\odot} \approx R/R_{\odot}$ for lower main sequence stars. 
\item While the post-processing described above substantially reduces the systematic variations in the light curves, some variations remain (see fig.~\ref{fig:RMSplot}). These variations give rise to increased power at low frequencies in the BLS periodogram of each light curve. As a result, long period transits should be treated with greater caution than short period transits. To account for this we subtract a mean filtered periodogram from the raw periodogram for each light curve before selecting the peak frequency. To mean filter the periodogram we replace each point in the periodogram by the mean value of the 200 points closest in frequency to the given point after applying an iterative $3\sigma$ clipping. Testing this modification on light curves with injected transits shows a very slight improvement ($< 1\%$) in the fraction of light curves for which the recovered period agrees with the injected period.
\end{enumerate}

We conduct both a high resolution and a low resolution BLS search on the light curves. The low resolution search is needed for the transit recovery simulations (\S~\ref{sec:Peps}) to finish in a reasonable amount of time. For the high resolution search we examine $20,000$ frequency values over a period range from $0.2$ to $5.0$ days using 200 phase bins at each trial frequency, using $R_{min} = 0.4R_{\odot}$ and $R_{max} = 1.3R_{\odot}$. For the low resolution search we examine 3072 frequency values over the same period range, using the same number of phase bins. In searching for candidates with the low resolution search we use identical parameters to those used in conducting the transit recovery simulations (\S~\ref{sec:Peps}). We use slightly different parameters for possible cluster members and for field stars (see the selections of these stars in \S~\ref{sec:Pmem} and \S~\ref{sec:deteff_fieldstars}). For candidate cluster members we take $R_{min} = 0.1R_{\odot}$ and $R_{max} = 1.1\times R_{phot}$ where $R_{phot}$ is the radius of the star estimated from its photometry assuming that it is a cluster member. For the field stars we take $R_{min} = 0.1 R_{\odot}$ and $R_{max} = 1.3 R_{\odot}$.  

As discussed by B06, a simple selection on the Signal Residue \citep[equation 5 in][]{Kovacs.02} or the Signal Detection Efficiency \citep[equation 6 in][]{Kovacs.02} generally yields a large number of false positive detections that must then be removed by eye. However, because it is very difficult to accurately include by eye selections in a Monte Carlo simulation of the transit selection process, it is necessary to devise automatic selection criteria that minimize the number of false positive detections to calculate a robust limit on the frequency of stars with planets. We discuss the three distinct selection criteria that we have devised in turn.

\subsubsection{Selection Criteria Set 1}\label{sec:selectcriteria1}

To devise the first set of selection criteria we inject fake transit signals into the light curves of 1366 stars that lie near the cluster main sequence on a CMD and choose criteria that maximize the selection of high signal-to-noise simulated transits, while minimizing the selection of false positives. The simulated transits have radii of $0.35$, $0.71$ and $1.0~R_{J}$, and inclinations of $90^{\circ}$. For each star/radius we simulate ten transits with periods ranging from $0.5$ to $5.0~{\rm days}$ and random phases. We use the relations between $r$-magnitude and mass, and $r$-magnitude and radius for the cluster that were determined in Paper I to estimate the mass and radius of each star. The light curves including injected transits are passed through the pipeline described in the previous subsection before running the BLS algorithm on them.

To calibrate the selection criteria we first must define the set of injected transits which we consider to be recoverable, we then adopt selection criteria that maximize the selection of these light curves while minimizing the selection of all other light curves. Note that we use information from the injected transits to define the recoverable sample, while we cannot use any of this information when defining the selection criteria. We consider the transit to be recoverable if $\Delta \chi^{2}_{BLS} - \Delta \chi^{2}_{BLS,0} < -100$, where $\Delta \chi^{2}_{BLS}$ is the reduction in $\chi^{2}$ for the best-fit BLS model to the light curve with the injected transit, and $\Delta \chi^{2}_{BLS,0}$ is for the light curve prior to injecting the transit. When a light curve satisfies this criterion the best-fit BLS model is strongly influenced by the transit signal. This can happen even if the identified period does not match the injected period, or a harmonic of the injected period, however, in these cases, detailed follow-up may eventually yield the correct period. Note that for transit selection criteria sets 2 and 3 we adopt the more conservative recoverability criterion that the recovered period must match the injected period, or one of its harmonics, to within $10\%$.

\begin{figure}[p]
\plotone{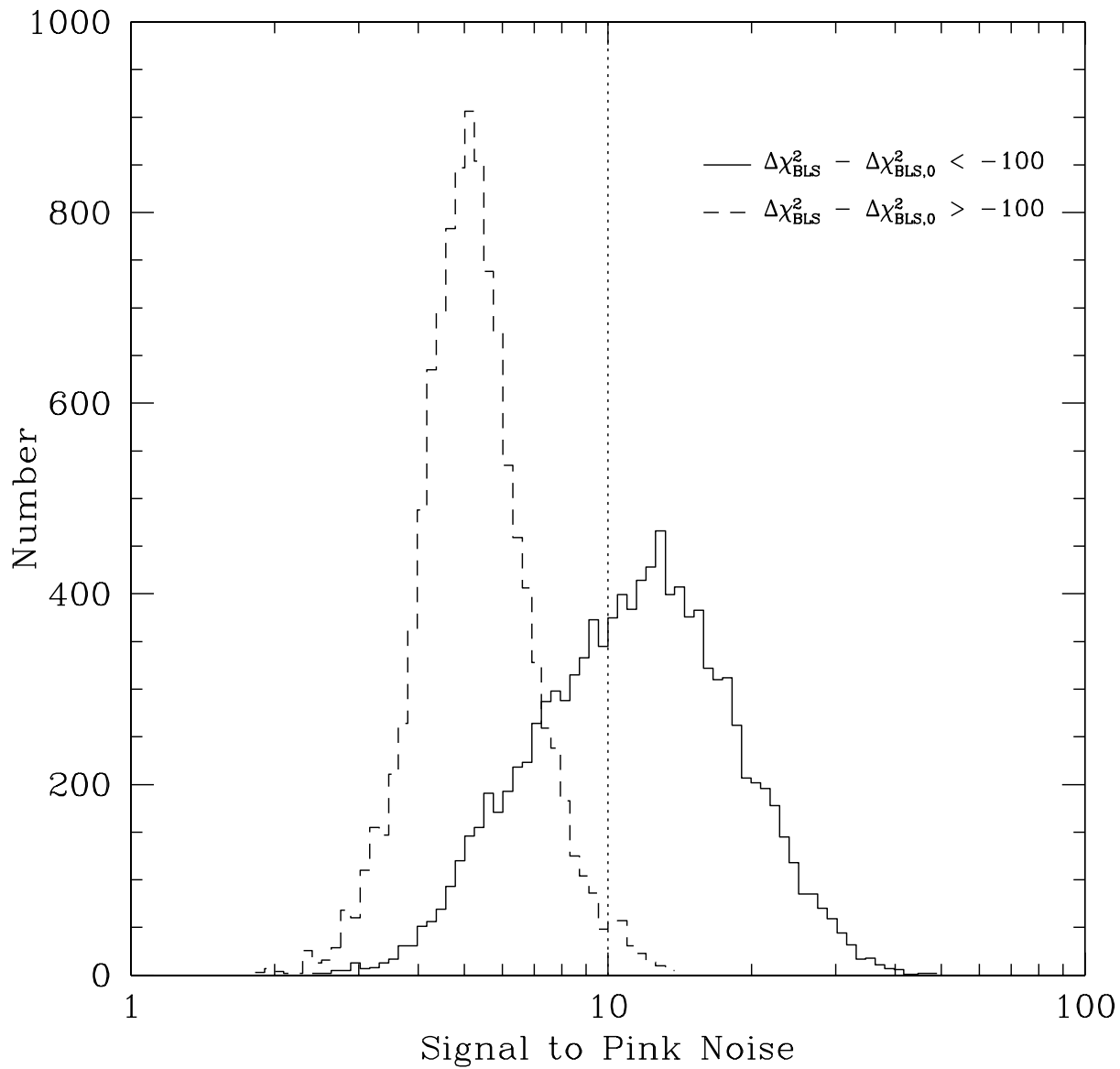}
\caption{Histograms of the signal pink noise ratio of light curves with injected transit signals. We show separate histograms for injected transits that should be recoverable ($\Delta \chi^{2}_{BLS} - \Delta \chi^{2}_{BLS,0} < -100$) and for those that may be unrecoverable. Light curves with $SN > 10.0$ are selected.}
%\caption{The ratio of the reduction in $\chi^{2}$ from the best-fit box-car transit model to the reduction in $\chi^{2}$ from the best-fit inverse transit model vs. the signal to pink noise ratio for a number of light curves with injected transit signals (left). We show the relation both for injected transits that should be recoverable ($\Delta \chi^{2}_{BLS} - \Delta \chi^{2}_{BLS,0} < -100$) and those that may be unrecoverable. (Right) the relation on the left is rotated (see the text for the definition of $X$ and $Y$), the dashed lines show the selection to distinguish between recoverable and unrecoverable transits.}
\label{fig:transselectcalib_1}
\end{figure}

\begin{figure}[p]
\plotone{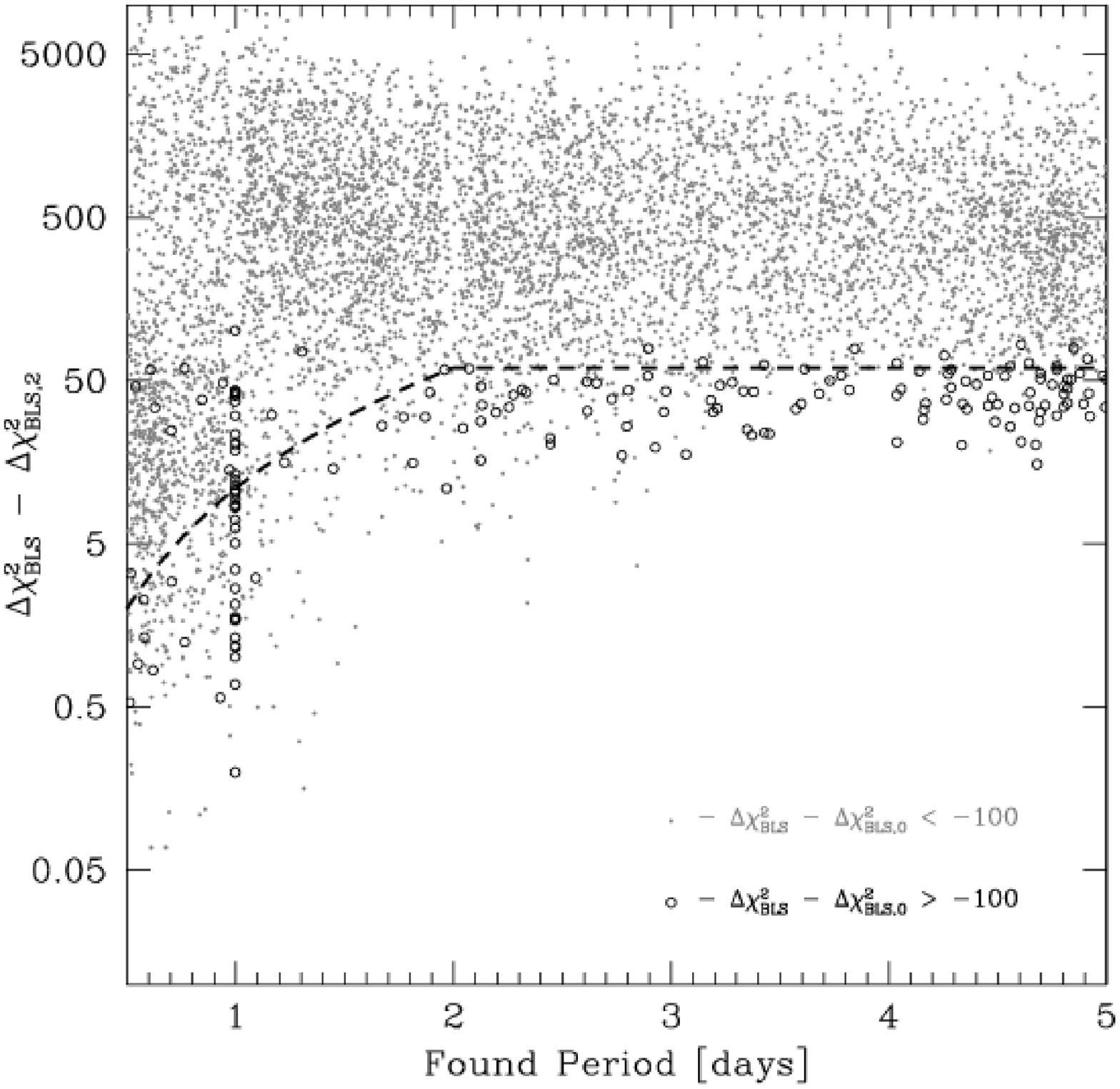}
\caption{The difference in $\chi^{2}$ reduction from the best-fit box-car transit model and the second best-fit model vs. the period of the best-fit model for light curves that pass the selection in figure~\ref{fig:transselectcalib_1}. Light curves below the dashed line are rejected.}
\label{fig:transselectcalib_3}
\end{figure}

We find that the following selection criteria accurately distinguishes between the recoverable and non-recoverable transits:
\begin{enumerate}
\item Following B06 we use BLS to identify both the best-fit transit signal and the best-fit inverse transit signal for each light curve. For the best-fit transit signal we calculate the signal-to-pink-noise ratio \citep{Pont.06} via:
\begin{equation}
SN^{2} = \frac{\delta^{2}}{\sigma_{w}^{2}/n_{t} + \sigma_{r}^{2}/N_{t}}
\label{eqn:sigtopink}
\end{equation}
where $\delta$ is the depth of the transit, $n_{t}$ is the number of points in the transit, $N_{t}$ is the number of distinct transits sampled, $\sigma_{w}$ is the white noise, and $\sigma_{r}$ is the red noise at the time-scale of the transit. To calculate the white noise for a light curve we subtract the best-fit BLS model from the light curve and set $\sigma_{w}$ equal to the standard deviation of the residual. To calculate the red noise we bin the residual light curve in time with a bin-size equal to the duration of the transit and set $\sigma_{r}$ equal to its standard deviation. Note that while this technique provides a convenient method to determine an individual red and white noise estimate for each light curve, it overestimates the noise for an uncorrelated signal by a factor of $\sim \sqrt{2}$. We select candidates that have $SN > 10.0$. Figure~\ref{fig:transselectcalib_1} shows this selection.
%We then define
%\begin{eqnarray}
%X & = & 0.986SN + 0.167\Delta \chi^{2}_{BLS}/\Delta \chi^{2}_{BLS,inv} \\
%Y & = & -0.167SN + 0.986\Delta \chi^{2}_{BLS}/\Delta \chi^{2}_{BLS,inv}
%\end{eqnarray}
%and select light curves with $X > 10$ or $Y > 1.5$. We also require $SN > 10.0$. Figure~\ref{fig:transselectcalib_1} shows these selections.
%\item Light curves that do not pass the above selection are then selected if they have $SN > 8$ and $\Delta \chi^{2}_{BLS} - \Delta \chi^{2}_{BLS,2} < -100$, where $\Delta \chi^{2}_{BLS,2}$ is the reduction in $\chi^{2}$ for the second highest peak in the BLS periodogram. Figure~\ref{fig:transselectcalib_2} shows this selection.
\item As shown in figure~\ref{fig:transselectcalib_1}, some of the transit signals that are not expected to be detectable pass the selection from the previous step. We can further reduce these potential false alarms with negligible loss of detectable transits by the following selection. Let $\Delta \chi^{2}_{BLS}$ refer to the reduction in $\chi^{2}$ for the best-fit transit signal and $\Delta \chi^{2}_{BLS,2}$ refer to the reduction in $\chi^{2}$ for the second non-inverse transit peak in the BLS spectrum. The light curves passing selection 1 are then selected if they have $P > 2.0~{\rm days}$ and $\Delta \chi^{2}_{BLS} - \Delta \chi^{2}_{BLS,2} < -60$ or $P \leq 2.0~{\rm days}$ and $\Delta \chi^{2}_{BLS} - \Delta \chi^{2}_{BLS,2} > -10.95P^{2.45}$ as shown in figure~\ref{fig:transselectcalib_3}. 
\end{enumerate}

%\begin{figure}[p]
%\plotone{transselectcalib_2.eps}
%\caption{The signal to pink noise ratio vs. the difference in $\chi^{2}$ reduction from the best-fit box-car transit model and the second best-fit model for light curves that do not pass the cut in figure~\ref{fig:transselectcalib_1}. Light curves that do not pass the previous selection will still be selected if they lie above and to the right of the dashed lines.}
%\label{fig:transselectcalib_2}
%\end{figure}

\begin{figure}[p]
\plotone{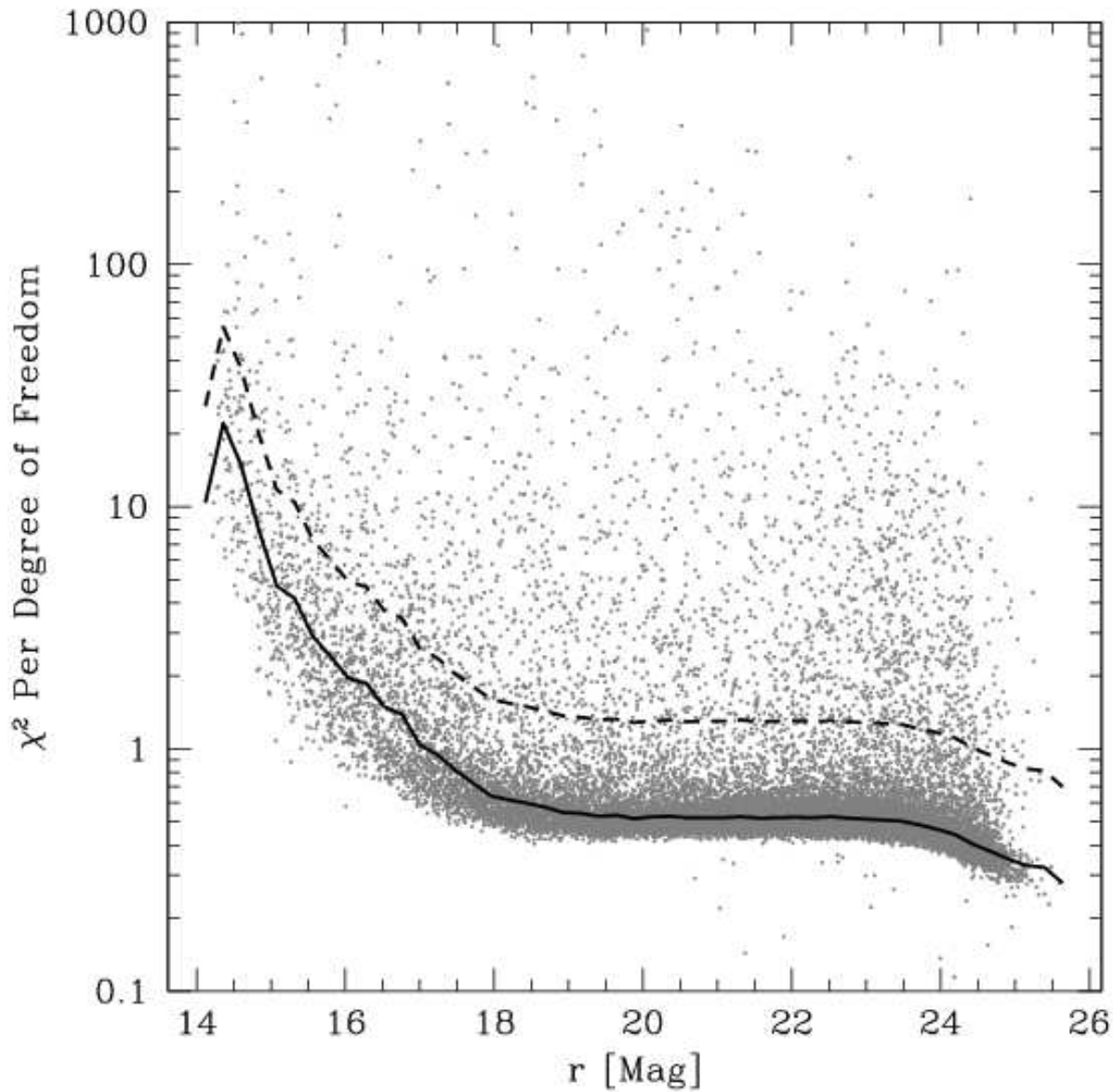}
\caption{$\chi^{2}$ per degree of freedom of the light curves without injected transits after subtracting the best-fit box-car transit model. The solid line shows the median $\chi^{2}$ per degree of freedom as a function of magnitude. Light curves above the dashed line ($2.5$ times the solid line) are rejected as transit candidates. See the text for a discussion of why $\chi^{2}$ per degree of freedom for stars between $18 \la r \la 23$ approaches $\sim 0.5$ rather than $\sim 1.0$.}
\label{fig:expectedchi2}
\end{figure}

When the above selections are applied to the actual data, more than 100 light curves are selected. Many of these are obvious false positives which can be eliminated by applying a few additional selections.
\begin{enumerate}
\addtocounter{enumi}{2}
\item We reject stars with fewer than 1000 points in their light curves, most of these are near saturation and show strong systematic variations even after applying TFA.
\item Many of the false positives are faint stars located near much brighter stars, these can be rejected by only considering stars with an average instrumental magnitude brighter than 17.0 (corresponding roughly to $r < 22.0$).
\item Other false positives come from very noisy light curves that can be rejected by requiring the standard deviation of the residual light curve after subtracting the best-fit box-car transit model be less than $0.1~{\rm mag}$.
\item Many false positives show an anomalous faint set of points that occur on only one night. Following B06 we require that the fraction of $\Delta \chi^{2}_{BLS}$ that comes from one night must be less than 0.8.
\item We reject light curves for which the best-fit box-car transit has a period between $0.99$ and $1.02~{\rm days}$ or less than $0.4~{\rm days}$.
\item Finally we require that $\chi^{2}$ per degree of freedom of the light curve residual after subtracting the best-fit box-car transit model must be less than $2.5$ times the expected $\chi^{2}$ per degree of freedom. The expected $\chi^{2}$ per degree of freedom is taken to be the median value as a function of magnitude for the full ensemble of light curves (figure~\ref{fig:expectedchi2}). Note that the values of $\chi^{2}$ per degree of freedom for stars between $18 \la r \la 23$ approach $\sim 0.5$ rather than $\sim 1.0$. This is due to a bug in our differential photometry routine whereby the formal differential flux uncertainties returned are not set to the flux scale of the reference image. We note that a similar bug is present in the differential photometry routine of the ISIS 2.1 package on which our photometry routine is based. The flux scale for the reference image that we use is set to that of a 30 second exposure taken in good seeing conditions. The formal errors are overestimated for longer exposures that were taken in poor seeing conditions. We identified this bug after completing most of the analysis presented in this paper, we do not, however, expect it to significantly affect the statistical results presented here. Note that since the photometry that we are using is dominated by red noise, rather than white noise, the formal photometric errors, which assume Gaussian, uncorrelated noise, do not accurately describe the uncertainties in the data. We have therefore used a cut on the signal-to-pink noise, which is determined empirically from the light curve, rather than relying on the formal uncertainties. The formal uncertainties do affect the relative weightings of points, modifying them may thus yield slight changes in the TFA fitting procedure as well as in the periods that are identified by BLS. If the formally correct weighting scheme had been adopted, the sample of selected candidates may have been slightly different, however, this is no different from making minor changes to the rather arbitrary selection criteria. For our statistical conclusions regarding the fraction of stars bearing planets, what is important is that we apply the same selections and weighting scheme to the transit injection simulations as we do to the actual data and that the post-selection vetting procedure that we apply to the actual data would not reject any of the injected candidates. Making a uniform change to the weighting scheme will affect the selection of real candidates and injected transits in the same manner, so the transit detection efficiency measured by the injection simulations will accurately reflect the detection efficiency for the survey.
\end{enumerate}

\subsubsection{Selection Criteria Set 2}\label{sec:selectcriteria2}

The above selection criteria were developed by manually inspecting the results of transit recovery simulations and defining cuts which appeared, by eye, to distinguish between successful recoveries and non-recoveries. While this method has the advantage that the criteria can be enumerated and easily visualized, it has the disadvantage that the cuts are fairly subjective and there is no guarantee that the selections optimally distinguish between recoveries and non-recoveries. 

To complement the above selection criteria, we have devised a set of selections using the Support Vector Machine (SVM) classification algorithm which has the advantage of being much less subjective, but the disadvantage of being difficult to visualize (\citealp{Vapnik.95}; a discussion of the algorithm can also be found in \citealp{Press.07} which we summarize here). This algorithm takes as input a set of training data which consists of $m$ points, $(\mathbf{x}_{i}, y_{i})$, where $\mathbf{x}_{i}$ is an $n$-dimensional vector of measureable parameters describing object $i$ (called features), and $y_{i}$ is either $+1$ or $-1$ and is used to divide the data into two classes. The algorithm then searches for a real-valued function $f(\mathbf{x})$ such that $f(\mathbf{x}_{i}) > 0$ for $y_{i} > 0$ and $f(\mathbf{x}_{i}) < 0$ for $y_{i} < 0$. The function is assumed to take the form:
\begin{equation}
f(\mathbf{x}) = \mathbf{W} \cdot \mathbf{\phi}(\mathbf{x}) + B
\end{equation}
where $\mathbf{\phi}(\mathbf{x})$ is a fixed $n$-to-$N$-dimensional transformation with $N$ typically larger than $n$, and $\mathbf{W}$ is an $N$-dimensional vector. To optimize $f$ one looks for a vector $\mathbf{W}$ that is normal to a hyperplane that separates the $\mathbf{\phi}(\mathbf{x}_{i})$ with $y_{i} = +1$ from the $\mathbf{\phi}(\mathbf{x}_{i})$ with $y_{i} = -1$ and that has a maximum perpendicular distance from the points nearest to it (called support vectors). In general, it is not always possible to find such a hyperplane, so instead one seeks to optimize $f$ by minimizing:
\begin{equation}
\frac{1}{2} \mathbf{W} \cdot \mathbf{W} + \lambda \sum_{i} \Xi_{i}
\end{equation}
subject to the constraint
\begin{eqnarray}
\Xi_{i} & \geq & 0, \\\nonumber
y_{i}f(\mathbf{x}_{i}) & \geq & \begin{array}{lr} 1 - \Xi_{i} & i=1,\ldots,m \end{array}
\end{eqnarray}
where $\Xi_{i}$ are free parameters which allow for discrepancies between the model and actual classes, and $\lambda$ is a fixed regularization parameter used to control the tradeoff between accurately classifying the training data and maximizing the perpendicular distance between the support vectors and the hyperplane. In practice, the problem is recast in Lagrangian formalism so that one specifies a kernel matrix with the property $K_{ij} = K(\mathbf{x}_{i},\mathbf{x}_{j}) = \mathbf{\phi(\mathbf{x}_{i})} \cdot \mathbf{\phi(\mathbf{x}_{j})}$ rather than the transformation $\mathbf{\phi(\mathbf{x})}$. 

To apply the SVM algorithm to our problem of devising transit selection criteria we use the \emph{SVMlight} package\footnote{C source code for \emph{SVMlight} is freely available at http://svmlight.joachims.org} \citep{Joachims.99, Joachims.02}. Before using the algorithm we apply a set of simple cuts which we found to be necessary to minimize the number of false positives:
\begin{enumerate}
\item Reject stars with an average instrumental magnitude fainter than 17.0 (corresponding roughly to $r > 22.0$).
\item Reject stars that have $RMS > 0.1~{\rm mag}$, where the $RMS$ here is the standard deviation of the residual light curve after subtracting the best-fit box-car transit model.
\item Reject stars with $SN < 10.0$, where $SN$ is the signal-to-pink noise given by eqn.~\ref{eqn:sigtopink}. Note for training the algorithm we use a less restrictive cut of $SN < 9.0$. 
\item Reject light curves for which the best-fit box-car transit has a period between $0.99$ and $1.02~{\rm days}$ or less than $0.4~{\rm days}$.
\end{enumerate} 
We train the algorithm on the simulated transit data described in \S~\ref{sec:Peps}. We use only $\sim 5000$ of the $\sim 250$ million simulated transits to allow the algorithm to converge in a reasonable amount of time. We take $y_{i} = +1$ for simulated transits that have $0.95 < P_{recover}/P_{inject} < 1.05$ where $P_{recover}$ and $P_{inject}$ are the recovered and injected transit periods respectively, and $y_{i} = -1$ for all other simulated transits. There are 15 features in the $\bold{x}_{i}$ vectors including: the recovered period, the fractional transit width, the transit depth, $SN$, the white noise, the red noise, $\Delta \chi^{2}_{BLS}$, $\Delta \chi^{2}_{BLS}/\Delta \chi^{2}_{BLS,inv}$, $(\Delta \chi^{2}_{BLS,2} - \Delta \chi^{2}_{BLS})/\Delta \chi^{2}_{BLS}$, the fraction of $\Delta \chi^{2}_{BLS}$ that comes from one night, the number of points in transit, the number of points observed less than $\tau$ minutes prior to transit ingress ($\tau$ is the duration of the transit), the number points observed less than $\tau$ minutes after transit egress, the number of distinct transits observed, and the ratio of $\chi^{2}$ per degree of freedom of the light curve residual after subtracting the best-fit box-car transit model to the expected $\chi^{2}$ per degree of freedom. When applying the algorithm we consider a transit to be recovered if the estimated value of $y$ is greater than $0.1$ as we found a number of false positives for which the algorithm returned values between $0.0$ and $0.1$.

\subsubsection{Selection Criteria Set 3}\label{sec:selectcriteria3}

As described in \S~\ref{sec:results}, when the first two sets of selection criteria are applied to transit recovery simulations the detection efficiency for $1.5R_{J}$ and larger planets drops unexpectedly for relatively bright stars. These large radius planets yield deep transit signals in the light curves of bright stars which are then distorted by steps 3-5 of the processing routines described in \S~\ref{sec:lcproc}. To recover these large radii planets we have devised a third set of selection criteria. For this set of criteria we run BLS on light curves processed through steps 1 and 2 of \S~\ref{sec:lcproc} (i.e. we do not remove one sidereal day period signals from the light curves, remove harmonic signals from the light curves, or apply TFA to the light curves). We then apply the SVM classification algorithm to the BLS results using the same training scheme and set of features as in \S~\ref{sec:selectcriteria2}. We train only on $1.0$, $1.5$ and $2.0R_{J}$ simulations.

\section{Transit Candidates}\label{sec:transcand}

Out of a total of 10,899 light curves for sources that were detected in $g$, $r$ and $i$, have an average instrumental magnitude brighter than 17 ($r \la 22$) and more than 1000 points, we select 16 transit candidates. Table~\ref{tab:transitcandidates} lists the candidates, their coordinates, photometry and BLS parameters and the selection criteria/resolutions which selected them. Phased light curves for the candidates are displayed in figure~\ref{fig:transitcandlcs}. As we discuss below, 12 of these candidates can be rejected as eclipsing binary stars by their large transit depths or by noting that the primary and secondary transit depths are unequal. Two of the candidates detected on the unprocessed light curves result from artifacts in the data which are removed by the processing routines. One of the candidates is a blend with a nearby deep eclipsing binary. One candidate Jupiter-sized transiting planet remains.

As seen in figure~\ref{fig:transitcandCMD}, several of the candidates (80014, 110021, 120050, 160017, and 170100) lie close enough to the cluster main sequence on the $g-r$ and $g-i$ CMDs to be selected as potential cluster members in \S~\ref{sec:deteff_clustermembers}, however all of these candidates can be rejected as either eclipsing binaries or a blend with an eclipsing binary (see \S~\ref{sec:indtranscand}). We can therefore only place an upper limit on the planet occurrence frequency of cluster members.

\begin{figure}[p]
\epsscale{0.7}
\plotone{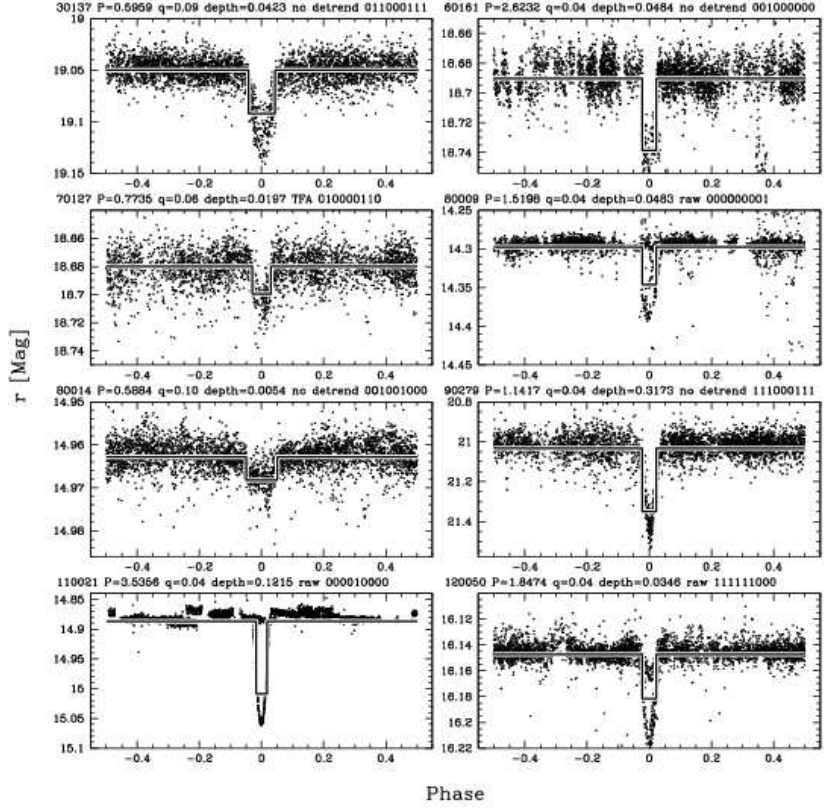}
\caption{Phased light curves for 16 candidate transiting planets selected by the pipeline discussed in \S~\ref{sec:transselectsubsection}. See \S~\ref{sec:indtranscand} for a discussion of each candidate. The ID is taken from the photometric catalog presented in Paper I. We list for each candidate the period in days, the fractional transit width $q$ and the transit depth in magnitudes returned by BLS. We note whether the displayed light curve has been processed through the full pipeline (TFA), through steps 1 and 2 only of \S~\ref{sec:lcproc} (no detrend), or if no processing has been applied (raw). We show raw light curves for a handful of candidates where $\sigma$-clipping removes some of the in-transit points. We also display a 9-bit flag to indicate which selection criteria sets and resolutions selected the candidate. From the left, the first three bits indicate if the candidate was selected by selection criteria sets 1, 2 and 3 respectively, using a high resolution BLS search. The middle three bits are for the low resolution BLS search applied to candidate cluster members, while the last three bits are for the low resolution BLS search applied to field stars.}
\label{fig:transitcandlcs}
\end{figure}

\addtocounter{figure}{-1}

\begin{figure}[p]
\epsscale{1.0}
\plotone{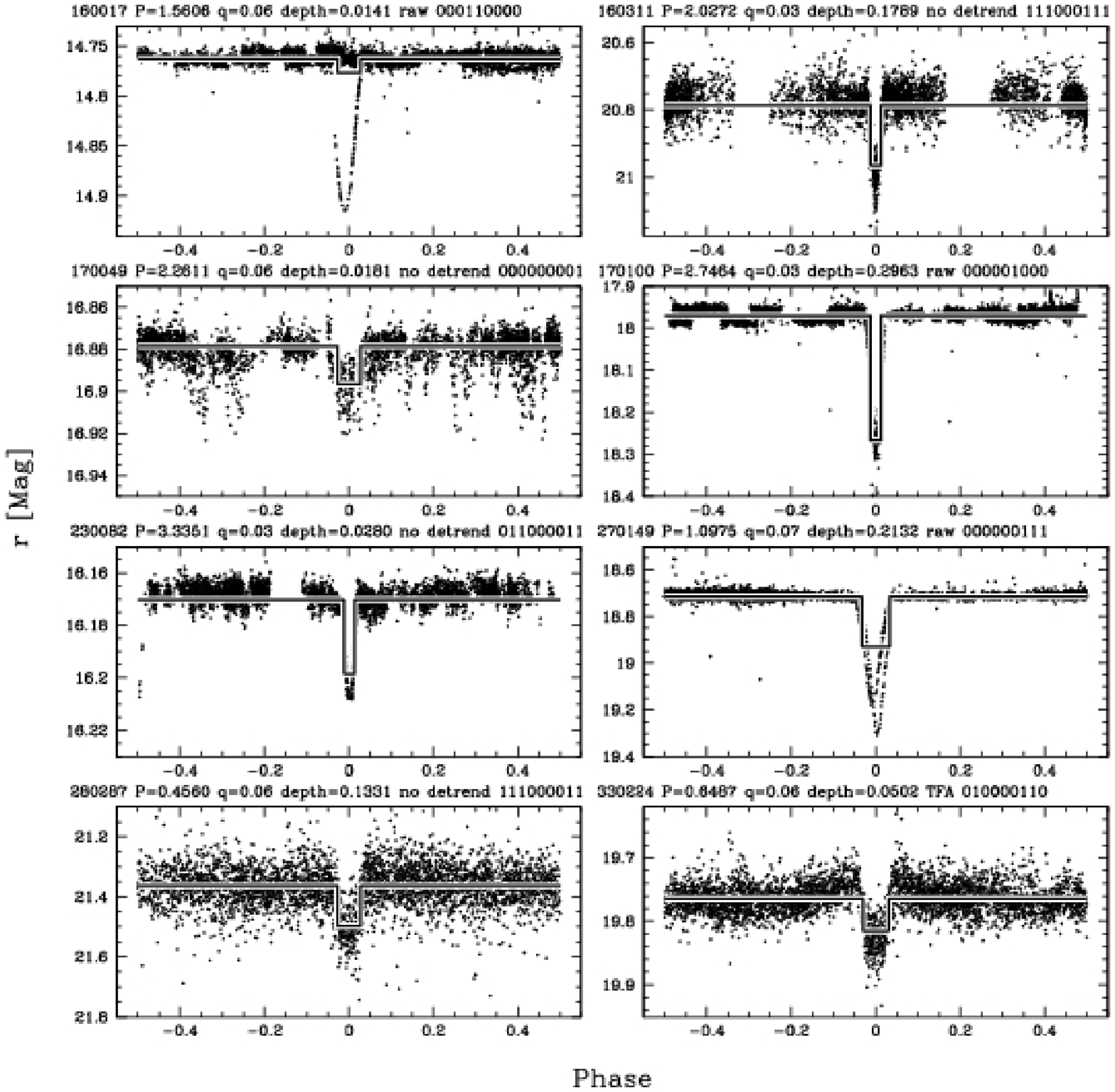}
\caption{continued.}
\end{figure}

\begin{figure}[p]
\plotone{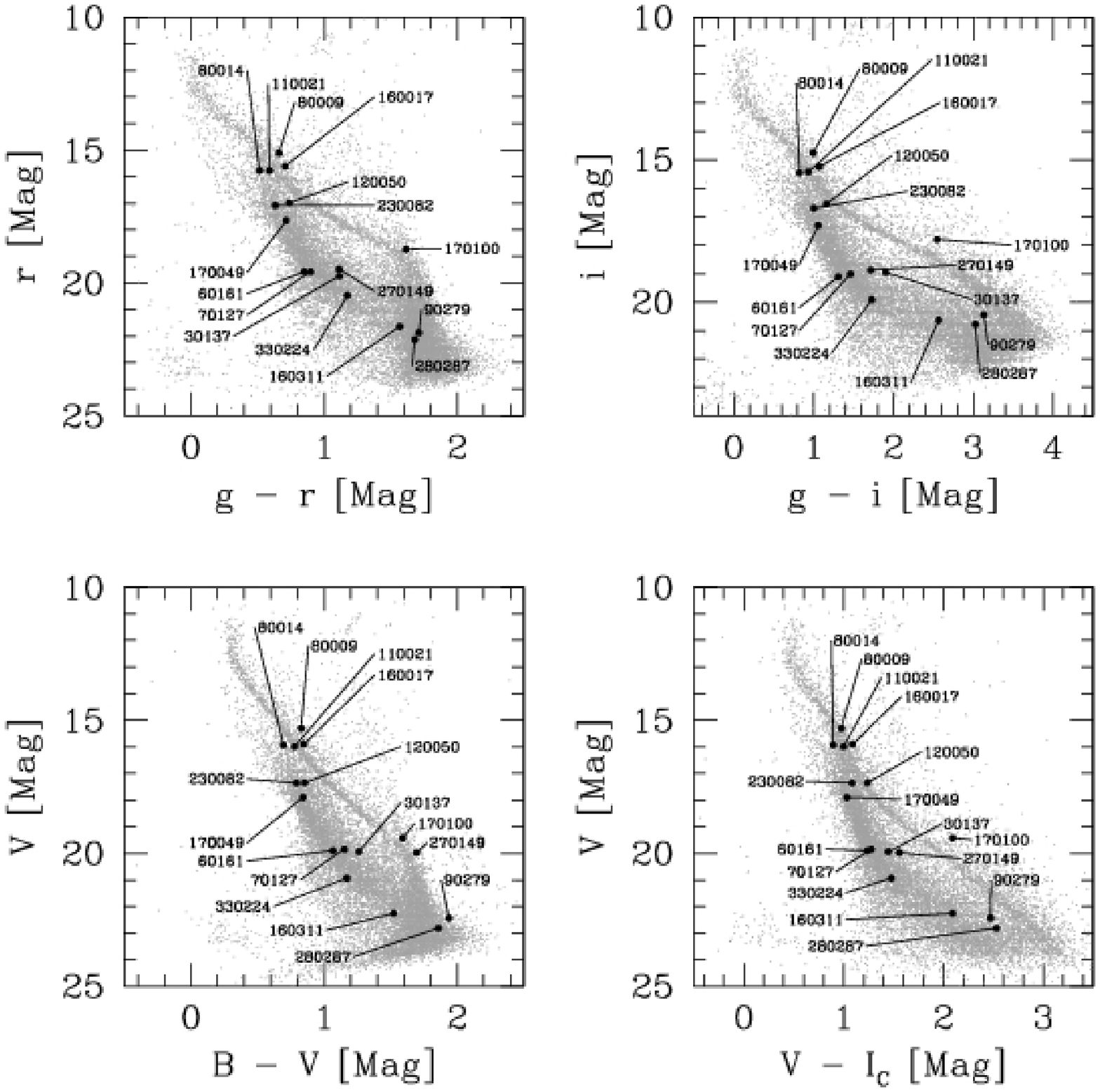}
\caption{The location of 16 candidate transiting planets on $g-r$, $g-i$, $B-V$ and $V-I_{C}$ CMDs. The light points show all stars, the dark points show the transit candidates. Five of the candidates lie near enough to the cluster main sequence on the $g-r$ and $g-i$ CMDs to be selected as potential cluster members (see figure~\ref{fig:CMDselection}).}
\label{fig:transitcandCMD}
\end{figure}

\subsection{Discussion of Individual Transit Candidates}\label{sec:indtranscand}

\begin{enumerate}
%\item[]\emph{10044} -- The raw light curve for this source shows a starspot variation with a period of $1.2617~{\rm days}$. This source is V1311 in the catalog of variables presented in Paper II. The transit feature that is identified has a fairly low signal to pink noise of $5.0$, also note a second transit-like feature at a phase of $-0.45$ in figure~\ref{fig:transitcandlcs}. We consider this to be a dubious candidate, the feature is likely a residual from not adequately modeling the starspot variation.
%\item[]\emph{30028} -- Examining the raw light curve for this candidate reveals that it is an eclipsing binary with a primary eclipse depth greater than $0.25$ mag. Our light curve samples only part of the primary egress, and many of these points were clipped during step 1 of the light curve processing described in \S~\ref{sec:lcproc}. We reject this candidate.
\item[]\emph{30137} -- This source is an eclipsing binary as evidenced by the secondary eclipse having an unequal depth to the primary eclipse when phased at twice the period shown in figure~\ref{fig:transitcandlcs}. This source is V1187 in the catalog of variables presented in Paper II. We reject this candidate.
%\item[]\emph{60017} -- The raw light curve shows significant scatter (greater than $0.01~{\rm mag}$) for an $r \sim 16.4~{\rm mag}$ source. This scatter, which is present in many of the light curves for sources on Chip 6, is significantly reduced by TFA. Cross-correlation analysis of Hectochelle spectra for this source (see Paper I) yields an effective temperature of $5850 \pm 145~{\rm K}$. Assuming a solar-mass star, the light curve has a stellar eta parameter \citep{Tingley.05} of $\eta_{\star} = 1.3$ which means that the transit depth and width are marginally inconsistent with the dip being due to a planet (light curves with $\eta_{\star} \leq 1$ are consistent). A transit depth of $1.4~{\rm mmag}$ for a $\sim 1.0 R_{\odot}$ star implies a planet radius of $\sim 0.3 R_{J}$. While not entirely compelling, this may be a promising candidate.
\item[]\emph{60161} -- The transit-like variation in this light curve is completely removed by TFA so this source only passes selection criteria set 3. The transit feature is due primarily to 2 nights. The images on these nights show a gradient in the background counts near the corner of Chip 6 where this star is located. The sense of the gradient is that points are fainter in the corner than near the center of the image. The use of a single master flat field image for the entire run appears to have failed for these two nights, the result is that many of the stars in the corner of Chip 6 show dips in their raw light curves on the two nights in question. This transit feature therefore appears to be spurious, and we reject it as such. We note that this is the only candidate that was selected by the high resolution search but not by the low resolution search.
%\item[]\emph{70093} -- The raw light curve shows a starspot variation with a period of $3.674~{\rm days}$. This source is V1452 in the catalog of variables presented in Paper II. The transit feature does not appear to be reliably repeated on multiple nights. There are 2 nights which contribute to the putative transit, one has tremendous scatter ($\sim 0.05~{\rm mag}$), the other is essentially the transit. The fraction of $\Delta \chi^{2}$ that comes from one night is $0.77$. We therefore consider this to be a dubious candidate.
\item[]\emph{70127} -- This source is a promising transit candidate around a field star. Cross-correlation analysis of the Hectochelle spectrum for the source yields $T_{eff} \sim 5000~{\rm K}$ though the uncertainty is very high due to the faintness of the source. Note that there is a nearby source to 70127 that may also contaminate this spectrum. Combining the $B$, $V$ and $I_{C}$ photometry for the source, we estimate that $(B-V)_{0} \sim 1.0$, and $E(B-V) \sim 0.16$ so that the star has a radius of $R_{\star} \sim 0.75 R_{\odot}$ and a mass of $M_{\star} \sim 0.8 M_{\odot}$. The light curve has \citet{Tingley.05} parameters of $\eta_{p} = 0.7$ and $\eta_{\star} = 0.7$, so it appears to be consistent with the transiting planet hypothesis. The source is located less than $2\arcsec$ from the edge of the chip, this appears to add some scatter to the light curve that is correlated with the seeing and removed by TFA. Also note that there is a $1.625$ mag fainter source, 70181, that is located only $0\farcs8$ from 70127. The fainter source also shows a dip in its light curve, however the dip in flux appears to be greater for 70127 than 70181, and the centroid of 70181 on the residual image appears to shift slightly in phase with the transit, whereas 70127 does not. Both of these factors suggest that 70127 is the real variable. The 'shoulder' just before transit is an artifact of post-processing the fairly high S/N light curve and does not appear in the unprocessed light curve. Fitting a \citet{Mandel.02} transit model to the light curve with quadratic limb darkening coefficients fixed to $u_{1} = 0.54$ and $u_{2} = 0.2$ from \citet{Claret.04}, which are appropriate for a $\sim 5000~{\rm K}$ dwarf star in the $r$ filter, and fixing $a/R_{\star} = 4.38$ assuming the stellar mass, radius and the orbital period of $0.77353~{\rm days}$, we find $R_{P} \sim 1.0 R_{J}$ and $\sin i \sim 0.99$. The expected RV amplitude for the star would be $K \sim 220~{\rm m/s}~(M_{P}/M_{J})$, where $M_{P}$ is the mass of the planet. To rule out astrophysical false positives, such as an M-dwarf secondary, would require additional follow-up spectroscopy which would be challenging given the faintness of the source ($r \sim 19.6$, $V = 19.85$). Note that \citet{Sahu.06} achieved a formal RV precision of $\sim 200~{\rm m/s}$ for their candidate SWEEPS-11, which has a comparable magnitude ($V=19.83$), so conducting spectroscopic follow-up for 70127 is not beyond the realm of possibility.
\item[]\emph{80009} -- This source appears to be an eclipsing binary, though there is not enough data to determine the period. There is a significant amount of scatter in the light curve that results from the source being nearly saturated. From the $B$, $V$, and $I_{C}$ photometry we estimate the reddening to the source is $E(B-V) \sim 0.18$ so that the source would have $(B-V)_{0} \sim 0.65$ and $M \sim 1.0M_{\odot}$. The primary eclipse depth appears to be deeper than the value returned by BLS (closer to $\sim 0.1~{\rm mag}$), the secondary source would have to be significantly larger than $2.0 R_{J}$ to create such a deep eclipse for this star. We reject this candidate.
\item[]\emph{80014} -- This shallow transit-like signal is the result of a blend with the $r \sim 18.3~{\rm mag}$ eclipsing binary, V29, that is located $\sim 3\arcsec$ away. We reject this candidate.
\item[]\emph{90279} -- This is an eclipsing binary with a period of $2.283~{\rm days}$. Phasing the light curve at twice the period shown in figure~\ref{fig:transitcandlcs} reveals that the secondary and primary eclipses are slightly unequal in depth. It is V1182 in the catalog of variables presented in Paper II. We reject this candidate.
%\item[]\emph{100043} -- While the phased light curve for this candidate is promising, we note that only transit ingresses have been observed. From Paper I we find a spectroscopic temperature of $6300 \pm 1660~{\rm K}$, which corresponds to a $\sim 1.2 M_{\odot}$ star (assuming a $1.0~{\rm Gyr}$, solar metallicity dwarf) and yields $\eta_{\star} \sim 1.1$. A transit depth of $2.6~{\rm mmag}$ for a $\sim 1.2 R_{\odot}$ star implies a planet radius of $\sim 0.5R_{J}$. Further photometric observations would be necessary to confirm that the observed dimming really corresponds to a transit.
\item[]\emph{110021} -- This source is a candidate cluster member with an estimated mass of $\sim 1.2 M_{\odot}$. The secondary object must be significantly larger than $2.0 R_{J}$ to yield an eclipse deeper than $0.1~{\rm mag}$, the source is thus an eclipsing binary. The source is V827 in the catalog of variables presented in Paper II. The period displayed in figure~\ref{fig:transitcandlcs} and listed in table~\ref{tab:transitcandidates} is the period returned by BLS. When phased at a period of $6.7667~{\rm days}$ the presence of a secondary eclipse becomes apparent, it is also apparent that the system has significant eccentricity, as well as an out of eclipse variation that phases with the proper orbital period. We reject this candidate.
\item[]\emph{120050} -- This is an eclipsing binary, note the dual eclipse depths at phase $0.0$. When phased at twice the period shown in figure~\ref{fig:transitcandlcs} it is clear that the light curve shows primary and secondary eclipses of differing depths. This source is V140 in the catalog of variables presented in Paper II. We reject this candidate.
%\item[]\emph{130027} -- The raw light curve for this source shows out of transit variability when phased at $1.337~{\rm days}$. This variability is removed by fitting a harmonic series to the light curve, the depth of the transit feature is also significantly reduced in this step. We consider this to be a dubious candidate.
%\item[]\emph{140051} -- 
%\item[]\emph{150079} -- The raw light curve for this source shows significant starspot variations at a period of 5.565 days. There appears to be some starspot evolution that is not adequately modeled by a simple harmonic series, the transit features in the TFA-cleaned light curve appear to be due to this. This source is V945 in the catalog of variables presented in Paper II. We reject this candidate.
\item[]\emph{160017} -- This source is an eclipsing binary. Its position on the CMD makes it a candidate cluster member with a mass of $\sim 1.2 M_{\odot}$. For such a primary the deep eclipse could not be caused by a planetary-sized companion. There is an out of eclipse variation that phases at $4.0989~{\rm days}$, however the eclipses do not phase at this period. There is not enough data to accurately determine the orbital period. In figure~\ref{fig:transitcandlcs} we show the raw light curve phased at the period returned by BLS for the light curve processed through TFA. The parameters listed in table~\ref{tab:transitcandidates} for this system are for the raw light curve, the S/N value is well below $10.0$ in this case. The $\sigma$-clipping routine clips much of the eclipse yielding a shallow transit that phases roughly at the period shown and yields a S/N value above $10.0$. This source is V791 in the catalog of variables presented in Paper II. We reject this candidate.
\item[]\emph{160311} -- This is an eclipsing binary, there is a shallow secondary eclipse with a depth of $\sim 0.05~{\rm mag}$ visible at phase $0.5$. This source is V716 in the catalog of variables presented in Paper II. We reject this candidate.
\item[]\emph{170049} -- The transit feature seen in the noisy unprocessed light curve is completely eliminated by TFA. The source lies within $2\arcsec$ of the edge of the frame and the variations in the light curve are strongly correlated with the image seeing. The transit feature is removed when the light curve is de-correlated against seeing. We reject this candidate as the variations appear to be spurious.
\item[]\emph{170100} -- This source is an eclipsing binary. When phased at twice the period displayed in figure~\ref{fig:transitcandlcs} it is clear that the primary and secondary eclipses are of unequal depth. The out of eclipse variations also phase at this period. This source is V1028 in the catalog of variables presented in Paper II. We reject this candidate.
%\item[]\emph{220020} -- 
\item[]\emph{230082} -- This source, V485, is most likely an F-M eclipsing binary. Note that when phased at the period recovered by BLS (as shown in figure~\ref{fig:transitcandlcs}) there appears to be a secondary eclipse at phase $-0.5$, when phased at a period of $1.6676~{\rm days}$, however, the putative secondary matches to the primary eclipse. Also note that the clipping procedure removes some points from the bottom of the eclipse, the eclipse in the raw light curve is $\sim 0.01~{\rm mag}$ deeper. From the $B$, $V$ and $I_{C}$ photometry we estimate $E(B-V) \sim 0.38$, so $(B-V)_{0} \sim 0.41$ which corresponds to a $\sim 1.5 M_{\odot}$, and $\sim 1.4 R_{\odot}$ primary star. For a star of this size, a $0.034$~mag transit requires that the companion have a radius greater than $2~R_{J}$.
\item[]\emph{270149} -- This source, V457, is an eclipsing binary. Note the deep, and distinct, primary and secondary eclipses. We plot the raw light curve for this star in figure~\ref{fig:transitcandlcs}, the clipping routine removed many of the points in eclipse which caused BLS to identify the wrong period and underestimate the depth.
\item[]\emph{280287} -- This is a grazing eclipsing binary. When the raw light curve is phased at twice the period found by BLS it is clear that the primary and secondary minima are of unequal depths. This source is V226 in the catalog of variables presented in Paper II. We reject this candidate.
%\item[]\emph{320051} -- The transit feature in this light curve is too wide for it to be due to a planet. The spectroscopic temperature of the source is $6620 \pm 350~{\rm K}$, which for a main sequence star, yields $M \sim 1.3 M_{\odot}$, $R \sim 1.4 R_{\odot}$. A planet transiting such a star with $P = 3.756~{\rm days}$ must have $q < 0.04$, and when we impose the stellar radius constraint in running BLS we find such a value for $q$ for this source. However, it is apparent that the transit feature is wider than this. When we do not impose a radius constraint we find $q = 0.15$ gives a best-fit for this period, the feature is therefore too wide for it to be due to a planet transiting a main sequence star. We reject this candidate.
\item[]\emph{330224} -- This source, V141 in the catalog of variables, is an eclipsing binary given its depth. The raw light curve reveals a strong out of eclipse variation with a peak-to-peak amplitude of $\sim 0.1$~mag that phases at the orbital period.
%\item[]\emph{350133} -- This source is located less than 10 pixels away from the edge of chip 35. The variations in the light curve are due to edge effects in the image subtraction routine as they correlate strongly with seeing. We reject this candidate.
\end{enumerate}

\section{Transit Detection Efficiency Calculation}\label{sec:deteff}

To calculate our planet detection efficiency we follow the procedure described by B06; we summarize the procedure here and perform the calculation separately for cluster members and field stars. 

\subsection{Cluster Members}\label{sec:deteff_clustermembers}

\begin{figure}[p]
\plotone{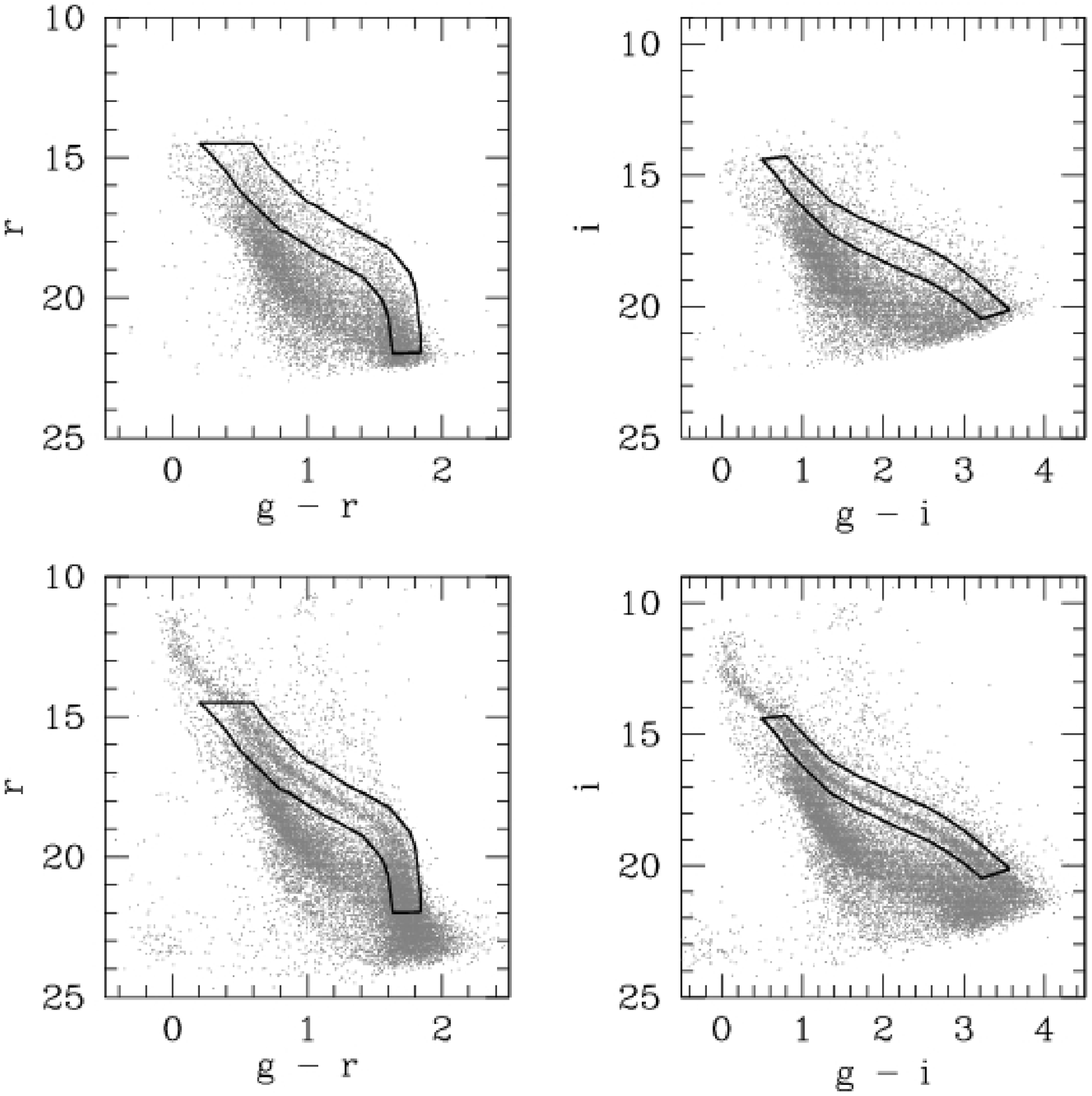}
\caption{The selection of stars near the cluster main sequence, which are used in computing the planet detection probability, are plotted on $g-r$ and $g-i$ CMDs. The bottom two panels show the CMDs for the field on the cluster, while the top two panels show the CMDs for a field adjacent to the cluster with the same Galactic latitude (see Paper I). The off-cluster field is used in determining the membership probability as a function of magnitude (fig.~\ref{fig:Memprob}). Stars falling within the solid black lines on both CMDs are selected as stars near the cluster main sequence.}
\label{fig:CMDselection}
\end{figure}

In figure~\ref{fig:CMDselection} we show the selection of stars near
the cluster main sequence on $g-r$ and $g-i$ CMDs. We select a total of 2475 stars that have an instrumental magnitude less than $17$, $RMS < 0.1~{\rm mag}$ and that have at least 1000 points in their light curves. Note that we expect $\sim 1450$ of these stars to be cluster members. We use these stars
to place a limit on the planet occurrence frequency for the cluster.
The total number of planets in the cluster that we expect to detect
with our survey is given by
\begin{equation}
N_{det} = f \sum_{i=1}^{N}P_{det,i}
\end{equation}
where $f$ is the planet occurrence frequency over a specified planet radius and orbital period range, $P_{det,i}$ is the probability of having detected such a planet for star $i$, and there are $N$ candidate cluster members in the survey. For a binomial distribution, the detection of no planets is inconsistent at the $\sim 95\%$ level when $N_{det} \ga 3$. The 95\% confidence upper limit on the planet occurrence frequency is then
\begin{equation}
f \leq 3/\left ( \sum_{i=1}^{N}P_{det,i} \right ).
\label{eqn:f}
\end{equation}
The planet detection probability for star $i$ is given by
\begin{equation}
P_{det,i} = \int \int \frac{d^{2}p}{dR_{p}dP}P_{\epsilon,i}(P,R_{p})P_{T,i}(P,R_{p})P_{mem,i}dR_{p}dP
\label{eqn:Pdet}
\end{equation}
where $P_{\epsilon,i}$ is the probability of detecting a transit of a planet with orbital period $P$ and radius $R_{p}$ for star $i$ if it has an orbital inclination that yields transits, $P_{T,i}$ is the probability that the planet has an inclination that yields transits, $P_{mem,i}$ is the probability that star $i$ is a cluster member, and $d^{2}p/dR_{p}dP$ is the joint probability distribution of $R_{p}$ and $P$.

There are four terms that contribute to equation~\ref{eqn:Pdet} that must be determined separately. Assuming random orientations, the term $P_{T,i}$ is given analytically by $P_{T} = (R_{\star} + R_{p})/a$, where $R_{\star}$ is the radius of the star and $a$ is the semi-major axis for an orbital period $P$ and stellar mass $M_{\star}$. For the term $P_{mem,i}$ we take the photometric membership probability of the star, which we calculate by comparing the CMD of the cluster to a CMD of a field off the cluster. The term $d^{2}p/dR_{p}dP$ must be treated as a prior, and the term $P_{\epsilon,i}$ is calculated via Monte Carlo simulation as discussed below.

\subsubsection{Calculating $P_{\epsilon,i}$}\label{sec:Peps}

We follow the procedure described by B06 to calculate $P_{\epsilon,i}$. This involves injecting limb-darkened transits into the light curves of potential cluster members and attempting to recover them. Transits are injected into the raw light curves, we then run each simulated light curve through the post-processing and transit selection routines described in \S~\ref{sec:transselect}. Note that we only conduct simulations on light curves that are not selected as transit candidates. Since there are only a few candidates, not including their efficiences should not change our results significantly.

To simulate transit light curves we use the \citet{Mandel.02} analytic model with $r$-band quadratic limb-darkening coefficients from \citet{Claret.04}. Note that we assume circular orbits for the short-period planets under consideration. We estimate the mass, radius and surface temperature of each star using the best-fit YREC isochrone \citep[][also see Paper I]{An.07}, we then estimate the limb-darkening coefficients for each star via linear interpolation within the \citet{Claret.04} grid, assuming $[M/H] = 0.045$ (Paper I) and $v_{turb} = 2~{\rm km/s}$.

To determine the dependence of $P_{\epsilon,i}$ on the orbital period we inject the transits in period bins ranging from $0.4$ to $5.0~{\rm days}$ with a logarithmic step size of $0.022$. For each period bin we inject $N_{trial}$ transits with periods distributed uniformly in logarithm over the bin, random phases and inclinations distributed uniformly in $\cos i$ over the range $0 \leq \cos i \leq (R_{\star} + R_{p})/a$. Following B06 we estimate that the error in the recovery fraction $f = N_{recover}/N_{trial}$ is given by
\begin{equation}
\sigma_{f} = \sqrt{f(1 - f)/N_{trial}}.
\end{equation}
For each period bin we initially adopt $N_{trial} = 20$ and continue simulating transits until $\sigma_{f} \leq 0.05$ using selection criteria set 1. We conduct simulations for planetary radii of 0.3, 0.35, 0.4, 0.45, 0.5, 0.6, 0.7, 0.8, 0.9, 1.0, 1.5 and $2.0 R_{J}$.

\subsubsection{Calculating $P_{mem,i}$}\label{sec:Pmem}

\begin{figure}[p]
\plotone{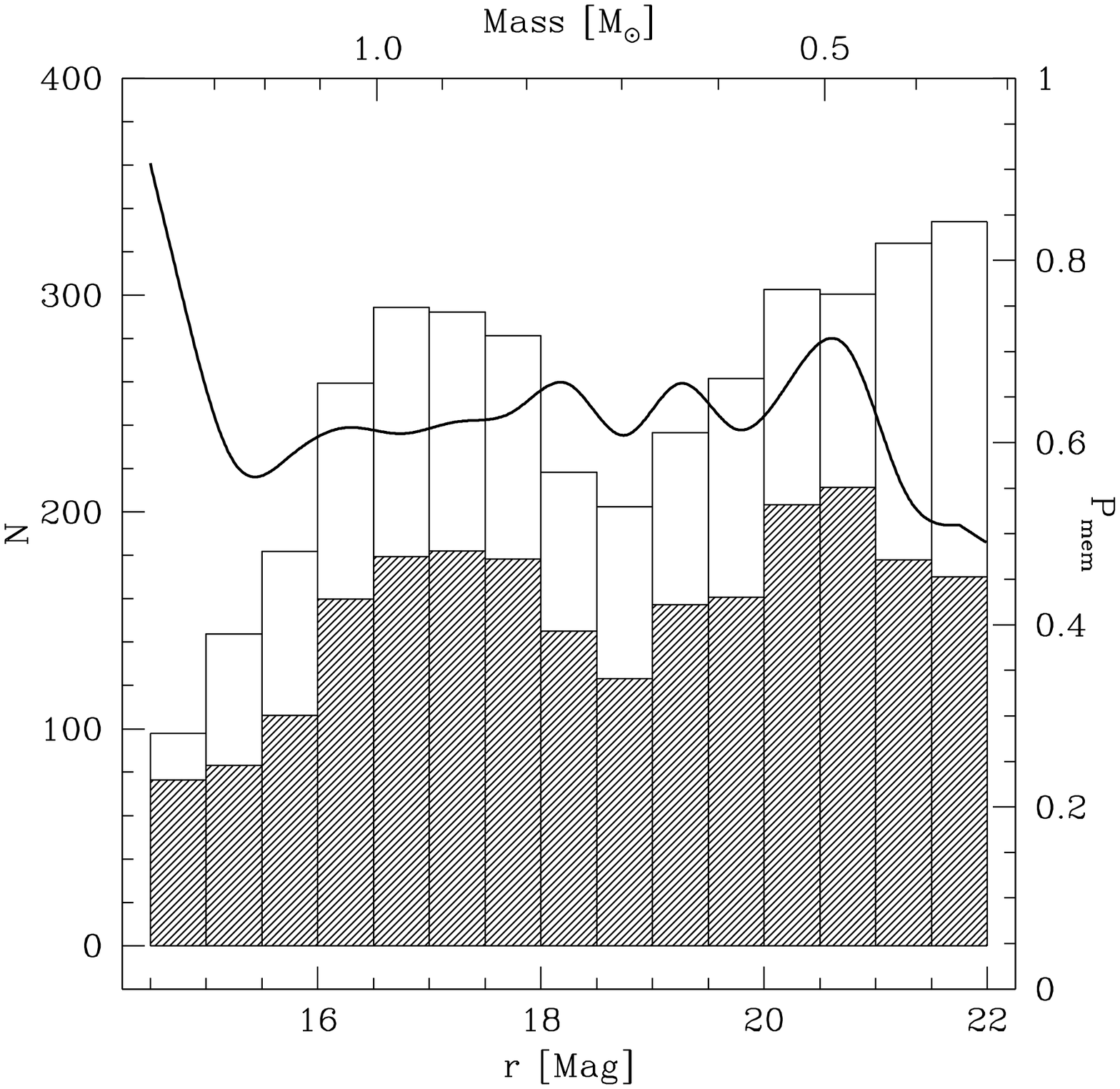}
\caption{Histogram of stars in the field of the cluster selected in figure~\ref{fig:CMDselection} as a function of $r$ magnitude (open histogram) compared to the histogram of cluster members (filled histogram) taken to be the difference between the histogram of the field on the cluster and the histogram of the field off the cluster. The solid line shows the membership probability (calculated as the ratio of the filled histogram to the open one) as a function of $r$. The histograms are taken from Paper I and include a correction for photometric incompleteness, but do not correct for spatial incompleteness. This figure is analogous to fig.~9 in B06.}
\label{fig:Memprob}
\end{figure}

To calculate the membership probability for each star we use the luminosity functions of the cluster and Galactic field along a strip in the $g-r$ and $g-i$ CMDs enclosing the cluster main sequence (see figure~\ref{fig:CMDselection}, the determination of the luminosity functions is described in Paper I). This gives the membership probability as a function of $r$-magnitude only. In figure~\ref{fig:Memprob} we show the membership probability as a function of $r$. Since this method ignores color information in assigning a probability to stars, it tends to give too high a probability to stars lying away from the cluster main sequence, and too low a probability to stars lying close to the main sequence. If the transit detection probability depends only on the $r$ magnitude of the star then this simplification should not bias the result. However, as shown in Paper II, cluster members have a higher probability of exhibiting photometric variability than field stars of the same magnitude, so we caution that it is possible that the transit detection probability may be lower on average for cluster members than for field stars. We consider this possibility in \S~\ref{sec:errors}.

\subsection{Field Stars}\label{sec:deteff_fieldstars}

To calculate the detection efficiency for field stars we follow a procedure that is similar to what we use for cluster members. In this case, however, the mass and radius of a star cannot be determined simply from the magnitude of the star. Instead we use Galactic models to estimate the mass and radius for each star. 

\begin{figure}[p]
\plotone{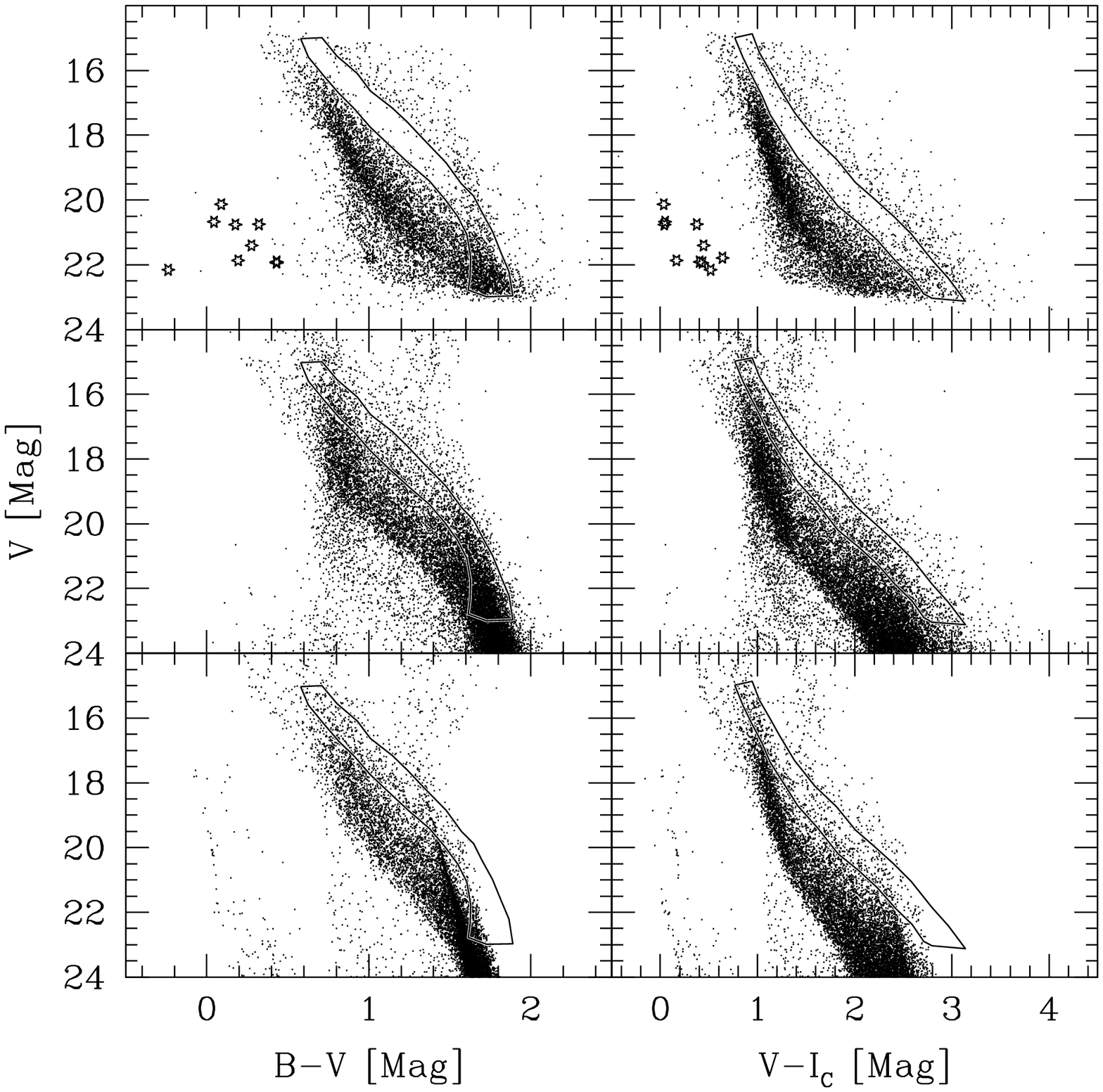}
\caption{$B-V$ (left) and $V-I_{C}$ (right) CMDs for stars selected as members of the Galactic field (top), for a simulated observation using the Besan\c{c}on Galactic models (middle) and for a simulated observation using the Trilegal Galactic model (bottom). For plotting purposes we have added Gaussian noise of $0.05~{\rm mag}$ to the $V-I_{C}$ colors from the Besan\c{c}on model. Stars not selected as cluster members that pass a number of cuts on magnitude, light curve RMS and the number of points in the light curve (see \S~\ref{sec:deteff_clustermembers}) are selected as members of the Galactic field. The solid lines show approximately the location of stars selected as potential cluster members, these stars are selected on $gri$ CMDs and are not included in the top two CMDs. To estimate the mass and radius for each star in the top two CMDs we take the values for the nearest star in the middle two CMDs. We use the Trilegal model to estimate the errors in the planet frequency upper limits that result from uncertainties in the Galactic model (\S~\ref{sec:errors}). Open stars in the top two plots show points that match to white dwarfs. We reject these stars from our sample.}
\label{fig:CMDselection_fieldstars}
\end{figure}

\begin{figure}[p]
\plotone{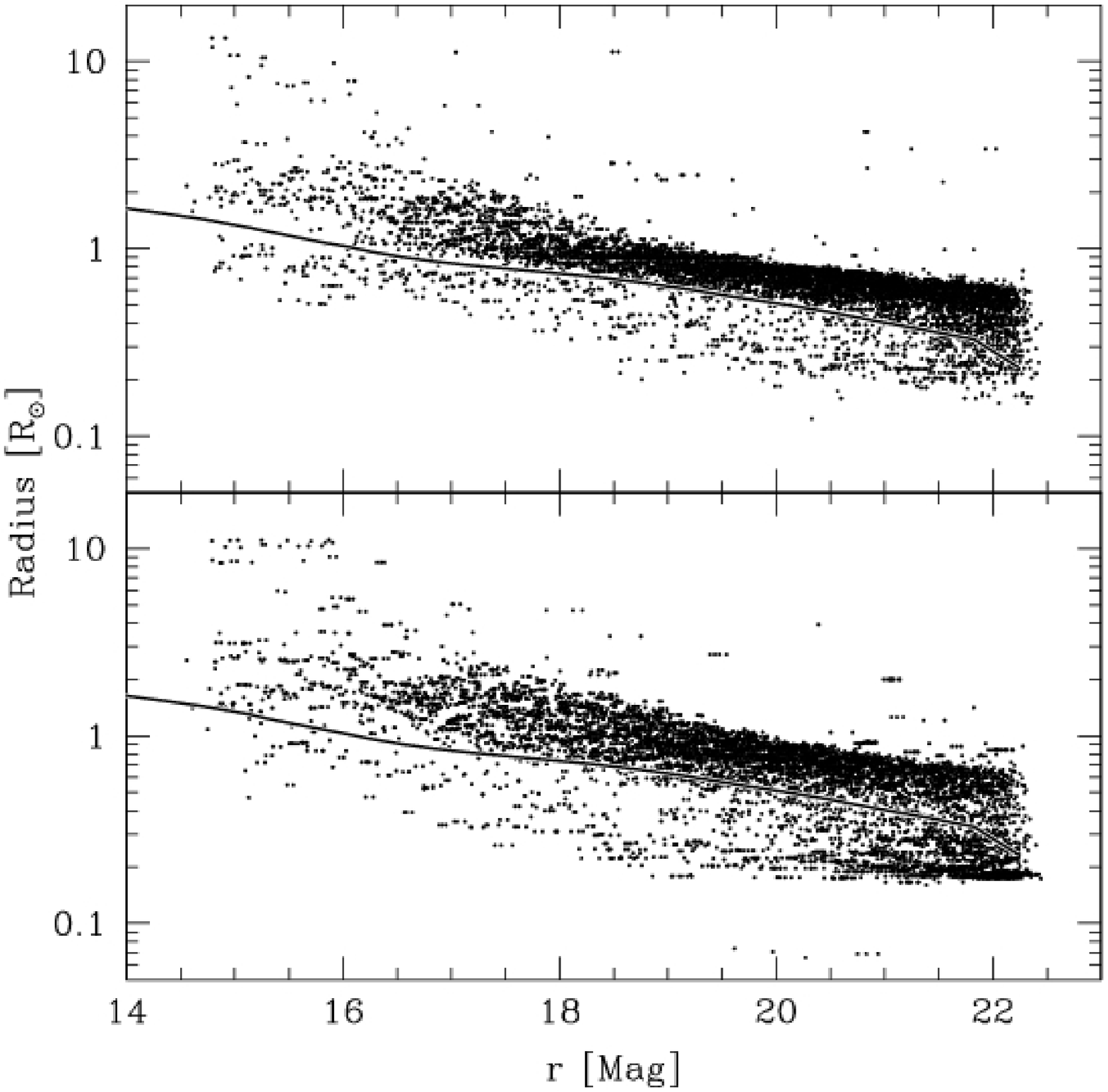}
\caption{The estimated stellar radius is plotted against the magnitude for Galactic field stars (solid points) when using the Besan\c{c}on Galactic model (top) and the Trilegal model (bottom). The solid line shows the relation for the cluster.}
\label{fig:Fieldstars_radius}
\end{figure}

First we select a sample of observed field stars that includes all stars with $g$, $r$, $i$, $B$ and $V$ photometry that have more than 1000 points in their light curves, have a light curve RMS that is less than $0.1$~mag, an instrumental $r$-band magnitude less than 17 (corresponding roughly to $r \la 22$) and were not selected as candidate cluster members. The sample of stars is shown on $B-V$ and $V-I_{C}$ CMDs in figure~\ref{fig:CMDselection_fieldstars}. A total of 7824 stars are selected in this fashion. We then obtain simulated photometric observations of a $24\arcmin \times 24\arcmin$ field centered at the Galactic latitude/longitude of the cluster using the Besan\c{c}on model \citep{Robin.03}. We assume an interstellar extinction of $0.7$ mag/kpc. We caution that this model is known to be unreliable along certain lines of sight, in particular for low Galactic latitudes \citep[e.g. see][]{Ibata.07}. For each observed star in our sample we choose the simulated star that minimizes
\begin{equation}
(V_{o} - V_{s})^{2} + (B_{o} - B_{s})^{2} + (I_{C,o} - I_{C,s})^{2}
\end{equation}
where the $o$ subscripts denote photometric measurements for observed stars, and the $s$ subscripts denote photometric measurements for simulated stars. We then assign the mass and radius of the simulated star to the observed star. We reject 10 stars which match to white dwarfs since these all lie in an isolated region of the CMD. The majority of the stars in the sample ($97\%$) match to dwarf stars ($\log g > 4.0$, cgs) rather than sub-giants or giants. Figure~\ref{fig:Fieldstars_radius} shows the estimated radii of the field stars as a function of magnitude compared with the values for the cluster. Note that at fixed magnitude, the majority of field stars have larger radii then the cluster stars. As a result, we expect the transit probability to be larger, but the overall S/N to be smaller at fixed magnitude for the field stars. Thus for signals for which the S/N is much larger than the threshold, the detection efficiency is larger for field stars, but for planets near the S/N threshold, the detection efficiency will be smaller. Thus the minimum detectable planet radius will be smaller for cluster stars, all else being equal. For the less common foreground field dwarfs, the opposite is true. 

We determine the detection efficiency by conducting transit injection and recovery simulations as described in \S~\ref{sec:Peps}. We conduct the simulations for 1000 randomly selected field stars. For each star we calculate $P_{det,i}$ using eq.~\ref{eqn:Pdet} where we now take $P_{mem,i} = 1$. The $95\%$ upper limit on the occurrence frequency, assuming no detections, can then be calculated using eq.~\ref{eqn:f} with the sum in the dominator multiplied by $7814/1000$ to scale from the simulated subset to the full sample.

\section{Results}\label{sec:results}

\begin{figure}[p]
\plotone{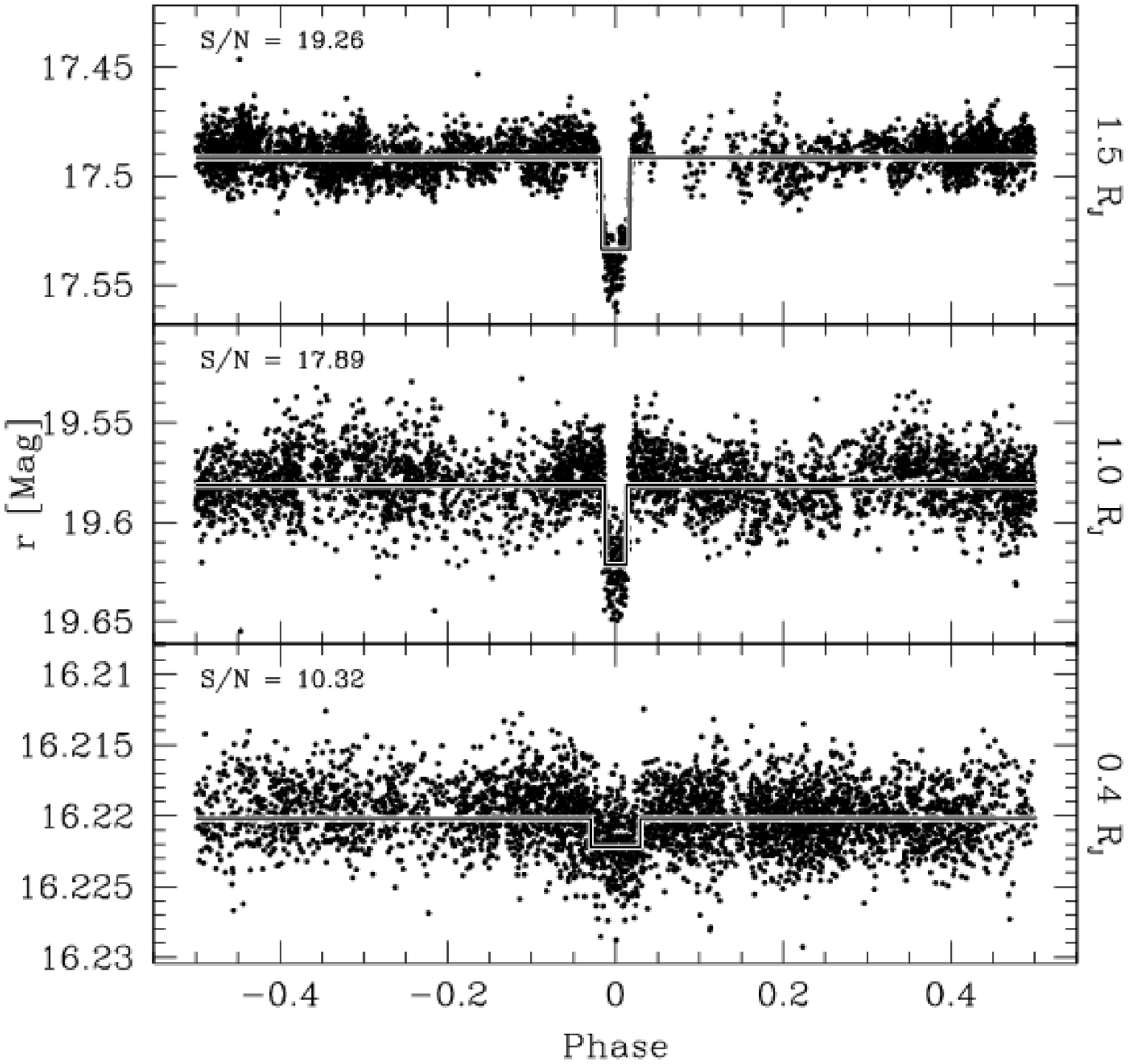}
\caption{Example phased light curves of simulated transits that are successfully recovered using selection criteria set 2 (\S~\ref{sec:selectcriteria2}). The simulated planets have radii of $1.5$, $1.0$, and $0.4~R_{J}$ while the stars have radii of $0.70$, $0.47$, and $0.83~R_{\odot}$ respectively. The injected periods are $1.77$, $1.44$ and $1.12~{\rm days}$ respectively. The line shows the best-fit box-car transit, we also list the signal-to-pink-noise ratio for each transit. Note that the examples shown correspond approximately to the median signal-to-pink-noise recoveries for each planet radius among all recovered simulations with periods between $1.0$ and $3.0~{\rm days}$. The $1.5$ and $1.0~R_{J}$ planets are easily recovered while the $0.4~R_{J}$ planet is marginally recovered.}
\label{fig:ExampleRecoveries}
\end{figure}

In figure~\ref{fig:ExampleRecoveries} we show several examples of transit injected light curves recovered in our simulations. Note that $1.5$ and $1.0~R_{J}$ planets are easily recovered. Also note that the S/N for $0.4 R_{J}$ planets is similar to that for $\sim 1.0R_{J}$ planets discovered by the field surveys \citep[see, e.g.,][]{Pont.06}.

\begin{figure}[p]
\plotone{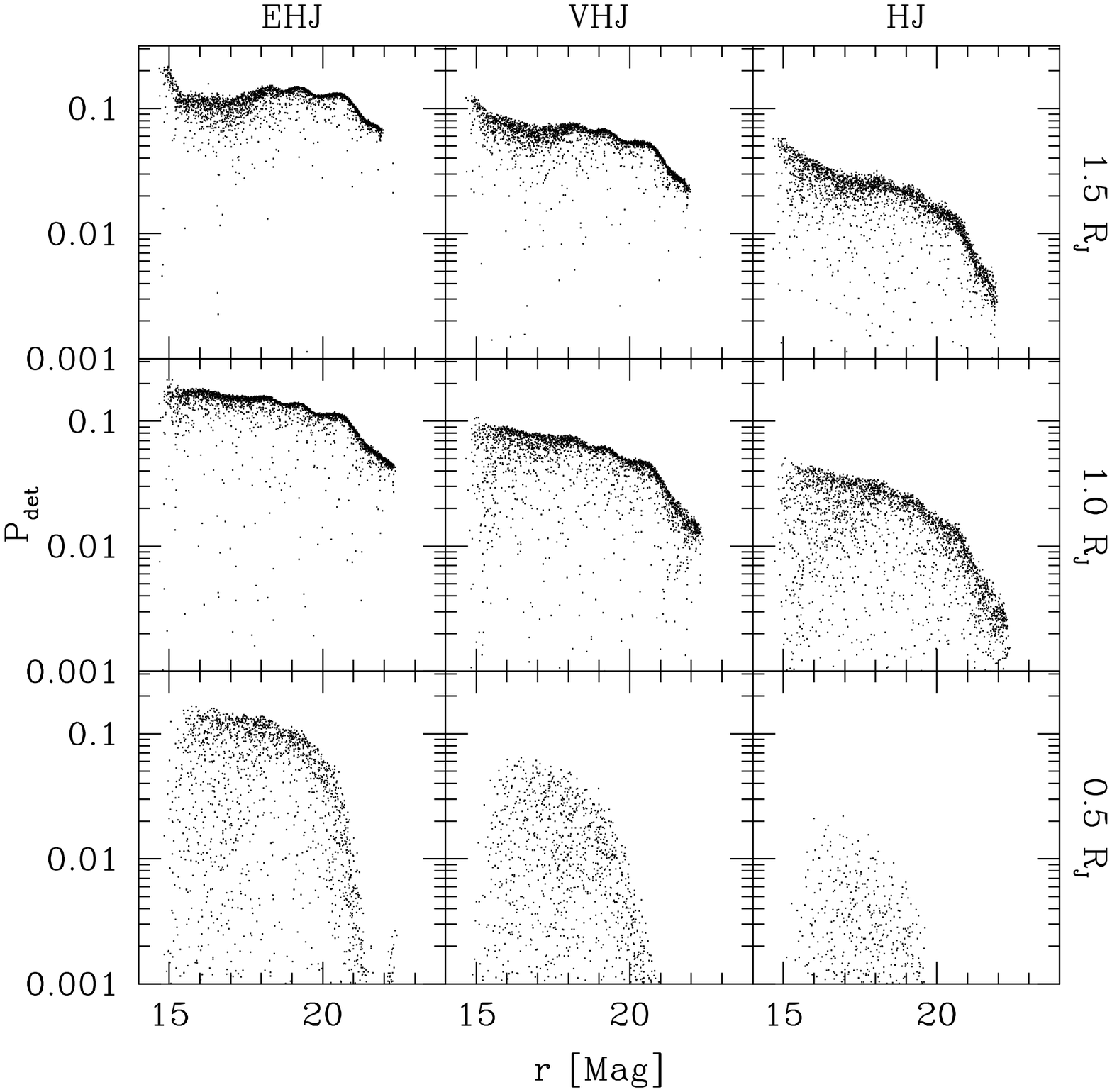}
\caption{The detection probability as a function of $r$-magnitude for transit injection/recovery simulations of candidate cluster members. The simulations are recovered using selection criteria set 2. We show the results for the HJ, VHJ and EHJ period ranges for several planetary radii. The detection probability decreases toward fainter magnitude since the transit probability at fixed period decreases for smaller stars. Note that for very high signal-to-noise transits ($1.5~R_{J}$) the detection probability is lower than expected as a result of the trend-filtering routines.}
%\caption{Same as figure~\ref{fig:DetectionProbvsMag1}, here we show the results for selection criteria set 2.}
\label{fig:DetectionProbvsMag2}
\end{figure}

In figure~\ref{fig:DetectionProbvsMag2} we plot the detection probability (eq.~\ref{eqn:Pdet}) as a function of magnitude for $0.5$, $1.0$ and $1.5R_{J}$ simulations on candidate cluster members recovered using selection criteria set 2. We show the results for Extremely Hot Jupiter (EHJ, $0.4 < P < 1.0~{\rm days}$), Very Hot Jupiter (VHJ, $1.0 < P < 3.0~{\rm days}$) and Hot Jupiter (HJ, $3.0 < P < 5.0~{\rm days}$) period ranges. For the HJ period range we use an upper limit of $5.0~{\rm days}$ rather than $9.0~{\rm days}$ since we do not attempt to recover planets with periods longer than $5.0~{\rm days}$. For each period range and planet radius we integrate eq.~\ref{eqn:Pdet} assuming a uniform probability distribution in $\log P$ and a Dirac-$\delta$ function distribution in $R$, i.e.
\begin{equation}
\frac{d^{2}p}{dP dR} \propto \delta(R - R_{0})/P
\end{equation}
where $R_{0}$ is the planet radius under consideration. Note that the detection probability for $1.5~R_{J}$ planets orbiting bright stars ($15.0 \la r \la 20.0$) drops relative to the probability for $1.0~R_{J}$ planets. There are two factors that contribute to this: steps 3-5 of the light curve processing routines described in \S~\ref{sec:lcproc} distort the high signal-to-noise transits for these large planets often leading to out-of-transit variations in the processed light curves, and the discrepancy between the limb-darkened transit signal and the box-car model becomes significant for the high signal-to-noise transits. When the light curve processing routines are not used (selection set 3) the drop in the detection probability of large planets is less significant, however the sensitivity to smaller planets is reduced.

\begin{figure}[p]
\plotone{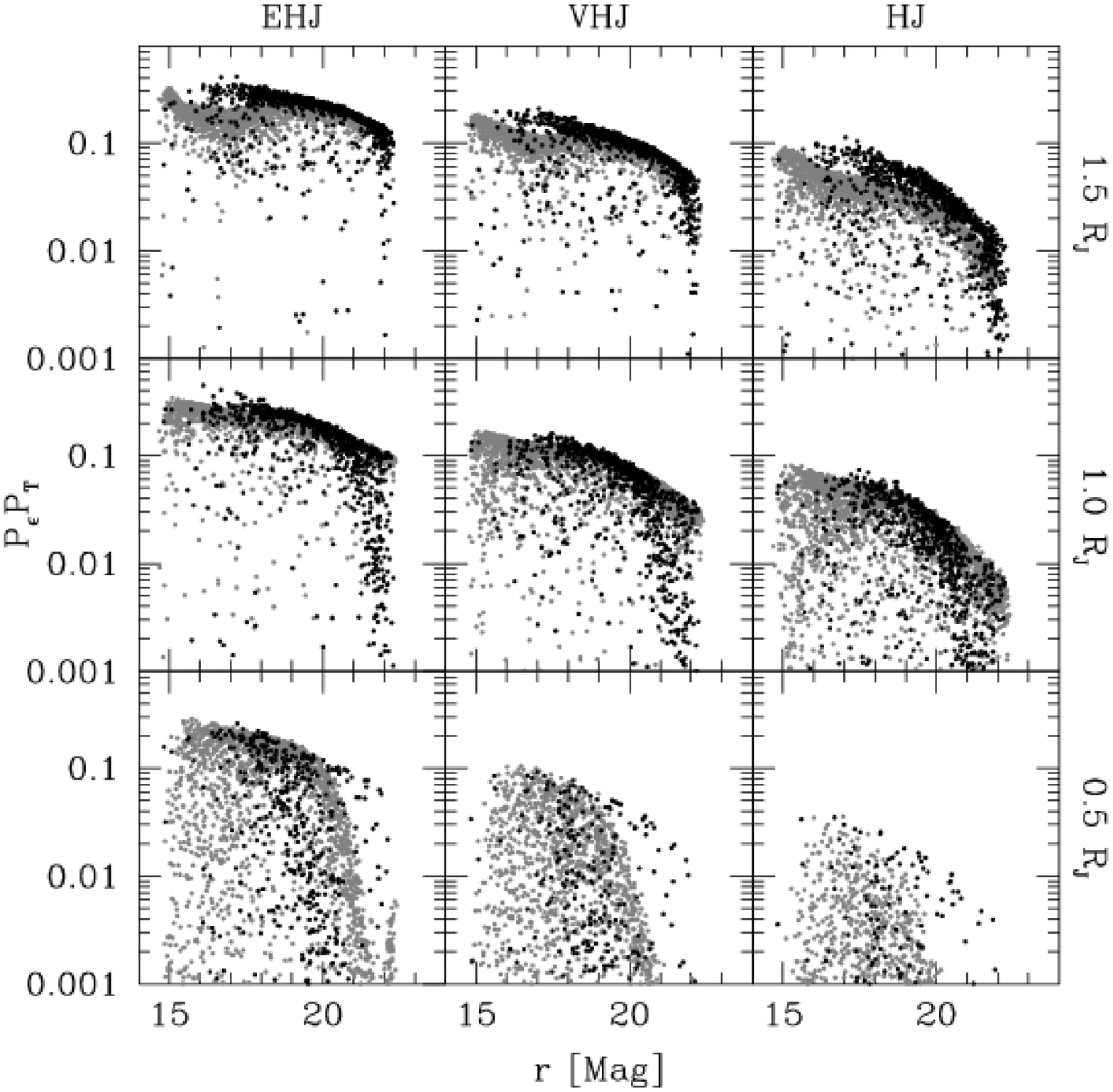}
\caption{The detection probability as a function of $r$-magnitude for field stars (dark points) compared with candidate cluster members (light points). We plot $P_{\epsilon}P_{T}$ which corresponds to $P_{det}$ when $P_{mem} = 1.0$ for cluster members and is equal to $P_{det}$ for field stars.}
\label{fig:PePTfieldstarsvsclustermembers}
\end{figure}

In figure~\ref{fig:PePTfieldstarsvsclustermembers} we compare the detection probability for field stars to the probability for cluster members, we set $P_{mem} = 1$ for the cluster members in making this comparison. As expected, the field stars generally show slightly higher detection probability at fixed magnitude than the cluster members for signals with S/N well above the detection threshold due to the higher transit probability, but smaller for those signals near the threshold due to the larger radii (and so shallower transits).

%\begin{figure}[p]
%\plotone{DetectionProbvsMag1.eps}
%\caption{The detection probability as a function of $r$-magnitude for transit injection/recovery simulations of candidate cluster members. The simulations are recovered using selection criteria set 1. We show the results for the HJ, VHJ and EHJ period ranges for several planetary radii. The detection probability decreases toward fainter magnitude since the transit probability at fixed period decreases for smaller stars. Note that for very high signal-to-noise transits ($1.5~R_{J}$) the detection probability is lower than expected as a result of the trend-filtering routines.}
%\label{fig:DetectionProbvsMag1}
%\end{figure}

%\begin{figure}[p]
%\plotone{DetectionProbvsMag3.eps}
%\caption{Same as figure~\ref{fig:DetectionProbvsMag1}, here we show the results for selection criteria set 3.}
%\label{fig:DetectionProbvsMag3}
%\end{figure}

\begin{figure}[p]
\plotone{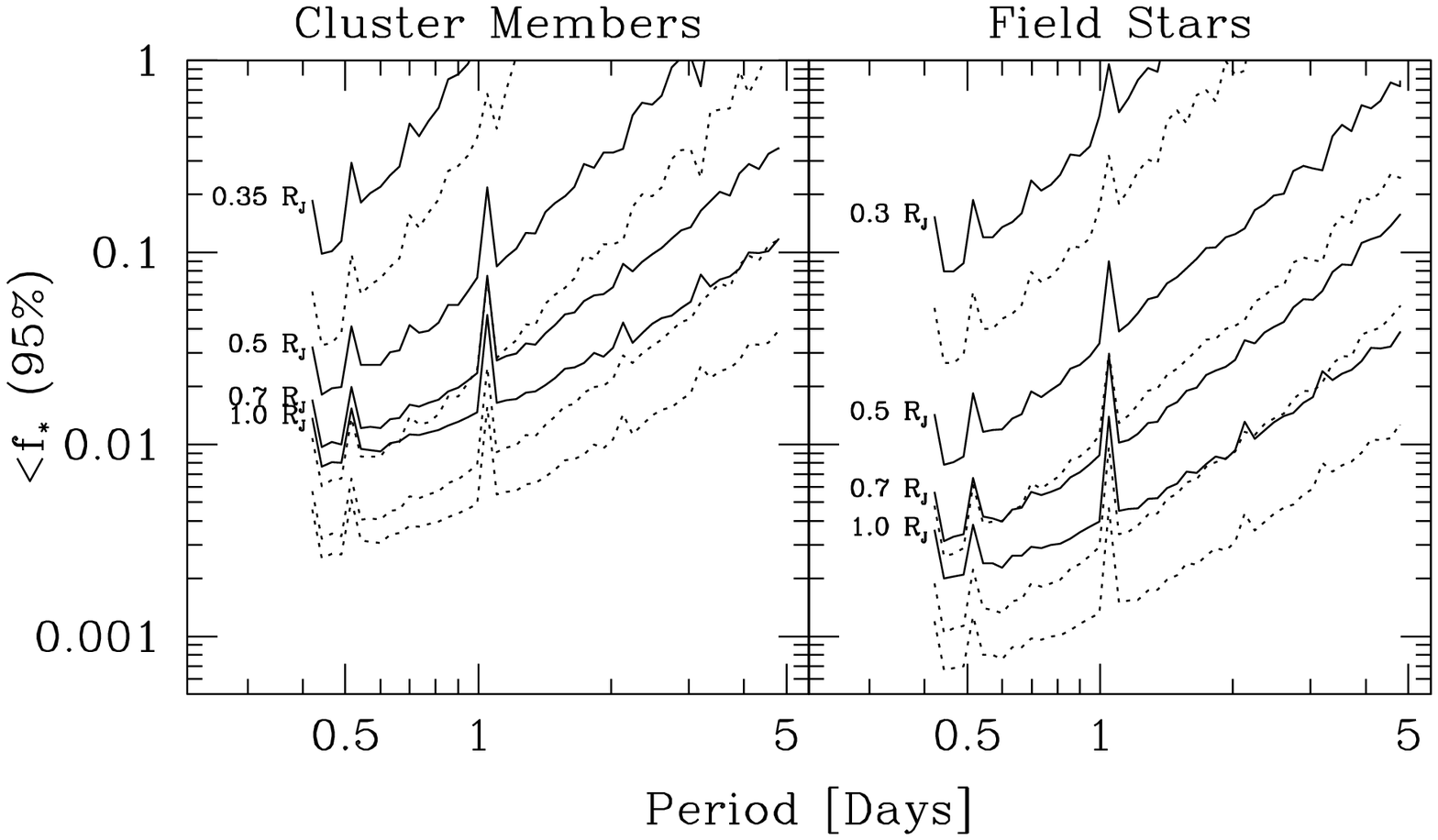}
\caption{The $\sim 95\%$ confidence upper limit (solid lines) on the fraction of cluster members (left) and field stars (right) with planets of a given radius as a function of orbital period. We show the results for $0.35$, $0.5$, $0.7$ and $1.0~R_{J}$ planets. For field stars we show $0.3~R_{J}$ rather than $0.35~R_{J}$. We also plot the $1-\sigma$ upper limits (dotted lines) for each of these planet radii.}
\label{fig:f95vsPeriod}
\end{figure}

\begin{figure}[p]
\plotone{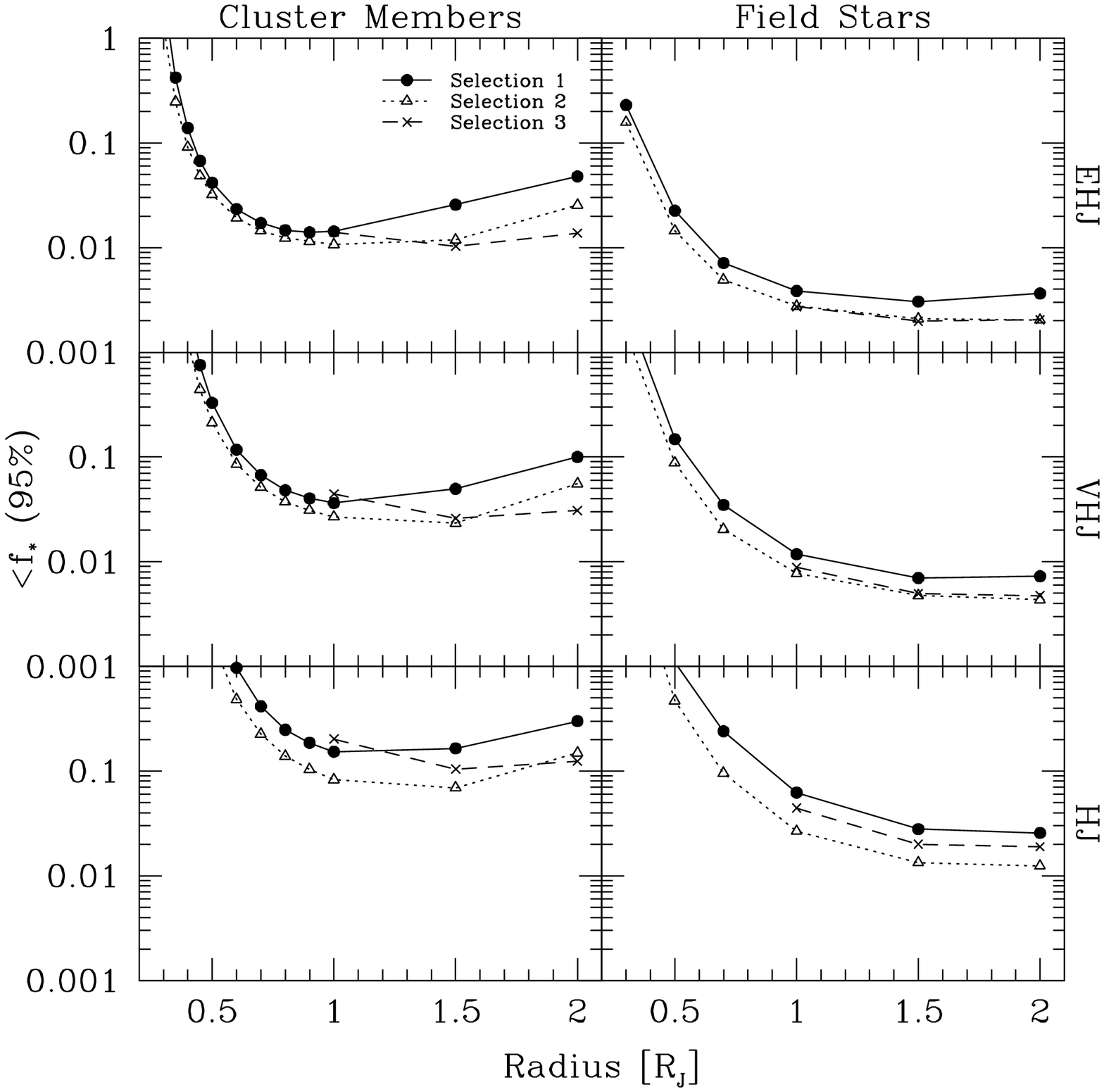}
\caption{The $\sim 95\%$ confidence upper limit on the fraction of cluster members (left) and field stars (right) with planets in a given period range as a function of planetary radius. We show the results for the EHJ ($0.4 < P < 1.0~{\rm days}$), VHJ ($1.0 < P < 3.0~{\rm days}$) and HJ ($3.0 < P < 5.0~{\rm days}$) ranges. We compare the results for each of the selection criteria sets discussed in \S~\ref{sec:transselectsubsection}.}
\label{fig:f95vsRadius}
\end{figure}

Figure~\ref{fig:f95vsPeriod} shows the $\sim 95\%$ and $1-\sigma$ confidence upper limits on the planet occurrence frequency for cluster members and field stars as a function of orbital period using selection criteria set 2. We assume the orbital periods are uniformly distributed in logarithm within logarithmic period bins of size $0.022$ and show the results for several different planet radii. In figure~\ref{fig:f95vsRadius} we show the $\sim 95\%$ confidence upper limits for cluster members as a function of planet radius for the EHJ, VHJ and HJ period ranges. We compare the results for the three distinct selection criteria sets. The results are listed as well in table~\ref{tab:f95vsRadius}. We find that the SVM-based selection (selections 2 and 3) outperforms the non-SVM-based selection (selection 1). The upper limits set using SVM are as much as 1.7 times smaller than the upper limits set using the non-SVM selection.

\begin{figure}[p]
\plotone{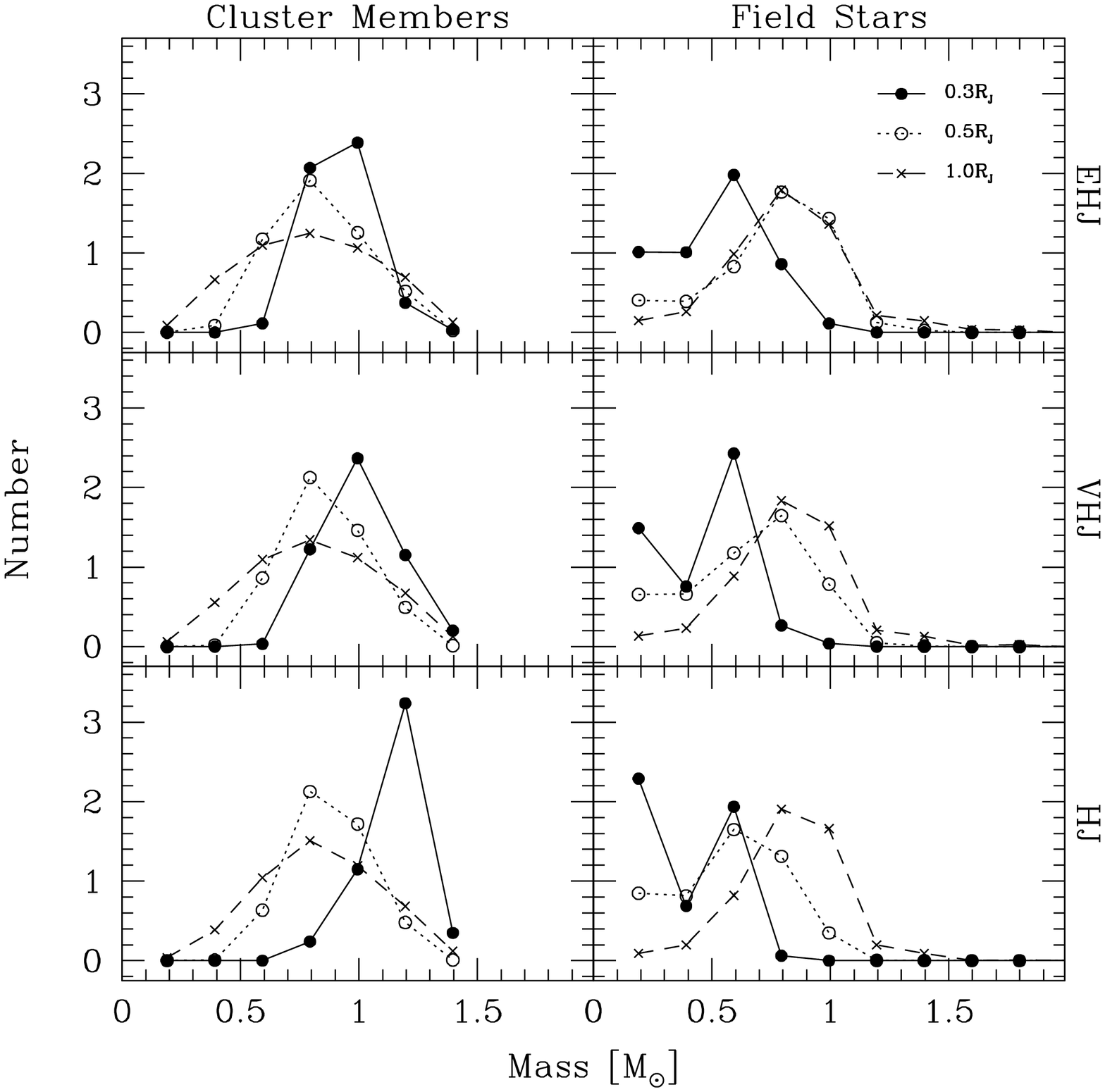}
\caption{The $P_{det}$ weighted distribution of stellar masses to which we have sensitivity to detect planets. The curves have been normalized to have unit integral over the range $0.09 M_{\odot} < M < 2.1 M_{\odot}$. The distributions are plotted for different planetary radii and period ranges, and are shown for cluster candidates (left) and field stars (right) separately. Note that for $0.3R_{J}$ planets the field star distribution is peaked toward smaller stars, whereas the cluster distribution is peaked toward larger stars. The small field stars to which we have sensitivity to $0.3R_{J}$ planets are foreground stars, at the distance of the cluster these stars are too faint for us to detect Neptune-sized planets around them. For larger planets the field star sensitivity distribution is peaked at around $0.8$-$1.0M_{\odot}$, whereas the cluster distribution is broader, but also peaked at around $0.8M_{\odot}$.}
\label{fig:DetectedStarHistogram}
\end{figure}

In figure~\ref{fig:DetectedStarHistogram} we show how the distribution of stellar masses to which we are sensitive to planets depends on planetary radius and period for the cluster candidates and field stars. Note that for $0.3R_{J}$ planets the field star distribution is peaked toward smaller stars, whereas the cluster distribution is peaked toward larger stars. The small field stars to which we have sensitivity to $0.3R_{J}$ planets are foreground stars, at the distance of the cluster these stars are too faint for us to detect Neptune-sized planets around them. For larger planets the field star sensitivity distribution is peaked around $0.8$-$1.0M_{\odot}$, whereas the cluster distribution is broader and peaked at around $0.8M_{\odot}$.

\subsection{Cluster Members}

For cluster members we place a $95\%$ confidence upper limit on the frequency of stars with Jupiter-sized EHJ, VHJ and HJ planets of $1.1\%$, $2.7\%$ and $8.3\%$ respectively. Note that all of these limits, as well as the ones we discuss below, come from selection criteria set 2. For smaller planets the limits rise. For the EHJ period range we can place a limit of $25\%$ on planets down to $0.35~R_{J}$, which is roughly the size of Neptune. For the VHJ period range, we can place a limit of $44\%$ on planets down to $0.45~R_{J}$, while for the HJ period range we can place a limit of $49\%$ on planets down to $0.6~R_{J}$.

\subsection{Field Stars}

The sample of field stars provides more stringent upper limits on the planet occurrence frequency than the sample of cluster members. For this sample we can place a $95\%$ confidence upper limit on the frequency of stars with Jupiter-sized EHJ, VHJ, and HJ planets of $0.3\%$, $0.8\%$ and $2.7\%$ respectively. This assumes that the candidate 70127 is not a real planet, if 70127 is real the frequency of $\sim 1.0~R_{J}$ planets in the EHJ period range would be $0.002\% < f < 0.5\%$ with $95\%$ confidence. In principle this frequency is small enough that planets in this period range could have escaped detection in most other radial velocity and transit surveys. Some transit surveys have probed enough stars to be sensitive to planets with these frequencies, as some of them have looked at as many or more stars than we have, but these surveys have generally not been very sensitive to hosts with masses as small as this host \citep[see][]{Gould.06,Beatty.08}. For the EHJ period range we can place a limit of $16\%$ on planets down to $0.3~R_{J}$, for the VHJ range a limit of $8.8\%$ on planets down to $0.5~R_{J}$ and for the HJ range a limit of $47\%$ on planets down to $0.5~R_{J}$. Extrapolating the curves in figure~\ref{fig:f95vsRadius}, we note that for the EHJ period range we do have some sensitivity even to planets as small as $\sim 2.5 R_{\oplus}$.

\subsection{Errors on the Upper Limits}\label{sec:errors}

There are a number of factors which may contribute errors to the upper limits. This includes uncertainties in $P_{\epsilon}$ from using a finite number of simulations, errors in $P_{mem}$ from using a finite sample, and systematic errors from blends and binaries, from errors in the assumed Galactic model and from the possibility that cluster members are more likely to be variable than non-cluster members. B06 give a detailed discussion of how most of these errors can be estimated. 

The fractional uncertainties in the $95\%$ upper limits for cluster members ($\sigma_{f_{<95\%}}/f_{<95\%}$) due to using a finite number of simulations to determine $P_{\epsilon}$ range from $\sim 1\%$ for $0.3 R_{J}$ simulations to $\sim 0.04\%$ for $1.0 R_{J}$ EHJ simulations. For field stars the fractional uncertainties are all less than $1\%$. These errors are negligible compared to other sources of uncertainty. 

Following B06 we note that the fractional uncertainty in the $95\%$ upper limit for candidate cluster members due to uncertainties in $P_{mem}$ is equal to the fractional uncertainty in the effective number of cluster members: $\sigma_{N_{\star,eff}}/N_{\star,eff}$, where $N_{\star,eff} = N_{\star}<P_{mem}>$.  We find $\sigma_{N_{\star,eff}}/N_{\star,eff} = 2\%$.

We expect binarity to be a more significant effect than chance alignments. As we noted in Paper II, we expect $\sim 1$ chance alignment within $0\farcs1$ of two point sources brighter than our detection threshold in our entire field. The effect of binarity on a transit survey is not straightforward. While the blending of light from two stars will dilute the transit signal, the primary star will be slightly smaller than what is assumed when injecting transits. Moreover, if the precision of the light curve is good enough that one could still detect the transit for a given planetary radius and primary star radius if the transit were more than a factor of 2 shallower, then such a planet would be detectable orbiting the primary star for any mass ratio and it may also be detectable orbiting the secondary star. In this case the total number of stars to which one is sensitive to planets is greater than estimated when binarity is neglected \citep[see][for further discussion about the effects of binarity]{Gould.06, Beatty.08}.  Binarity will only affect the detectability of a transit if the mass ratio is high, B06 estimate that the requirement is $q \ga 0.6$, which they argue is true for only $\sim 11\%$ of dwarf stars when the binary fraction $\sim 50\%$. There is an indication that the binary fraction along the main sequence in M37 is closer to $20\%$ \citep{Kalirai.04}, which would reduce the number of high mass ratio binaries to $\sim 4\%$ assuming that the mass ratio distribution in the cluster is the same as for the Galactic field. 

To estimate how uncertainties in the inferred masses and radii of field stars that results from uncertainties in the Galactic model affect the planetary frequency upper limits, we recompute the upper limits using the Trilegal 1.2 Galactic model \citep{Girardi.05}. We match the sample of field stars to a simulated set of photometric observations generated with the Trilegal model following the same procedure that we used in \S~\ref{sec:deteff_fieldstars} to match to the Besan\c{c}on model. The Trilegal model is shown in figures~\ref{fig:CMDselection_fieldstars} and~\ref{fig:Fieldstars_radius} and is generated assuming a local extinction law of $0.7$ mag/kpc. With this model we find that $96.5\%$ of stars in our field sample have $\log g > 4.0$, which is comparable to the dwarf fraction found with the Besan\c{c}on model. In \S~\ref{sec:deteff_fieldstars} we computed the transit detection probability for a sample of 1000 field stars by conducting transit injection/recovery simulations with stellar masses and radii adopted from the match to the Besan\c{c}on model. Here we avoid the expensive task of conducting additional transit injection/recovery simulations by matching stars in the Trilegal-matched sample to appropriate stars in the Besan\c{c}on-matched sample. We match each star, $i$, in the Trilegal-matched sample to the star, $j$, in the Besan\c{c}on-matched sample that minimizes
\begin{equation}
(0.2 \ln (10) (r_{i} - r_{j}))^{2} + (2(R_{i} - R_{j})/R_{i})^{2}
\end{equation}
where $r$ is the $r$-magnitude of the star, and $R$ is its radius. We then set the transit detection probability for star $i$ equal to the probability for star $j$ determined from the simulations. This choice of weighting between magnitude and radius minimizes differences between the expected transit signal to noise ratio of the two stars. The upper limits on the planet occurence frequency are then computed with eq.~\ref{eqn:f}. The results using selection criteria 2 are listed in table~\ref{tab:f95galmodel}. We find that the fractional difference in the planet frequency upper limits from the two Galactic models is $\la 10\%$. The Trilegal model yields a systematically lower upper limit for the $0.3 R_{J}$ and $0.5 R_{J}$ planets but a systematically higher upper limit for larger planets. In a similar manner we recalculate the upper limits using the Besan\c{c}on model generated with less extinction (0.5 mag/kpc). In this case we find that the upper limits for planets smaller than $1.0 R_{J}$ are systematically smaller while the upper limits for larger planets are systematically larger. The fractional differences in the upper limits for $0.7 R_{J}$ and smaller planets range from $\sim 5\%$ to $\sim 30\%$, while for larger planets they range from $\sim 2\%$ to $\sim 20\%$. We adopt a fractional uncertainty of $\sim 10\%$ on the upper limits due to uncertainties in the Galactic model.

To estimate the effect of variability on the upper limits for cluster members we compute the upper limits under the extreme case where all candidate cluster members that are variable have $P_{mem} = 1$ and other stars have $P_{mem} = (N(r)P_{mem,0}(r) - N_{var}(r))/(N(r) - N_{var}(r))$ where $P_{mem,0}$ is the value of $P_{mem}$ when variable and non-variable stars are weighted equally, $N(r)$ is the number of candidate cluster stars in magnitude bin $r$ and $N_{var}$ is the number of variable candidate cluster stars in magnitude bin $r$. The resulting $95\%$ upper limits for selection criteria set 2 are given in table~\ref{tab:f95vars}. For small radius planets the effect is significant, so that the upper limit on $0.35 R_{J}$ EHJ planets, for instance, increases to $38\%$ from $25\%$. For larger planets the effect of variability is less important, above $1.0R_{J}$ the fractional increase in the upper limit is less than $10\%$.

Assuming that binarity and variability only increase the upper limits, the fractional uncertainty on the upper limits for $1.0 R_{J}$ planets orbiting candidate cluster members is $\sim +11\%, -2\%$, for $0.35 R_{J}$ planets it is $\sim +50\%, -2\%$. For field stars we can neglect the uncertainty due to variability and the uncertainty on the membership probability, but we must include the uncertainty on the Galactic model, so that the fractional uncertainty on the upper limit for all planetary radii is $\sim +15\%, -10\%$.

\subsection{Comparison with Previous Results}

Our results can be compared both with previous results for transit surveys of open clusters, and with results from transit and RV surveys of the Galactic field. 

Our limits for the cluster on the frequency of EHJ, VHJ and HJ planets with $R = 1.5R_{J}$ of $1.2\%$, $2.3\%$ and $6.9\%$ respectively are substantially better than the corresponding limits for NGC 1245 by B06 of $1.5\%$, $6.4\%$ and $52\%$. The primary differences between the two surveys are that we observed approximately twice as many cluster members as B06 and we obtained significantly higher precision photometry at fixed stellar radius than B06. Note, however, that B06 use a period range of $3-9~{\rm days}$ for the HJ range whereas we use a range of $3-5~{\rm days}$. Since, by selection, we have zero sensitivity to planets with $5~{\rm days} < P < 9~{\rm days}$, extending our HJ range to the B06 range results in a limit of $\sim 15\%$ on the fraction of stars with planets of radius $1.5~R_{J}$ (assuming an underlying distribution in $P$ that is uniform in logarithm). \citet{Miller.08} have also conducted detailed Monte Carlo simulations of their open cluster transit survey and placed $95\%$ upper limits of $14\%$ and $45\%$ on VHJ and HJ with radii of $1.5R_{J}$. 

We can estimate the frequency of planets detected by RV surveys in the EHJ, VHJ and HJ period ranges using the results from \citet[][see figure 5 in their paper]{Cumming.08}. Out of their sample of 585 FGKM stars, there are 7 planets with $M > 0.1 M_{J}$ and $3~{\rm days} < P < 5~{\rm days}$, 1 planet with $M > 0.1 M_{J}$ and $1~{\rm day} < P < 3~{\rm days}$, and no planets detected with $P < 1~{\rm day}$. This yields $95\%$ confidence intervals for the occurrence frequencies of $0.48\% < f_{HJ} < 2.5\%$, $0.0043\% < f_{VHJ} < 0.95\%$, and $f_{EHJ} < 0.51\%$. Our results are not directly comparable to the RV survey results for two reasons: first, our limits are based on planetary radius while the RV survey limits are based on planetary mass, second, the distribution of stellar masses and metallicities is not the same for the two surveys. Nonetheless, if we assume all extra-solar giant planets have radii of $\sim 1.0R_{J}$ and that the planet occurrence frequency does not depend on stellar properties, then our upper limits of $f_{EHJ} < 0.3\%$, $f_{VHJ} < 0.8\%$ and $f_{HJ} < 2.7\%$ for field stars are consistent with the measured RV frequencies. 

\citet{Gould.06} used the results of the OGLE transit survey to determine the frequency of VHJ and HJ planets. Assuming planets follow a uniform distribution in radius between $1.0R_{J}$ and $1.25R_{J}$, they found $f_{VHJ} = 0.14^{+0.19}_{-.08}\%$ and $f_{HJ} = 0.31^{+0.35}_{-0.17}\%$ with $90\%$ confidence error bars. Furthermore, they placed a $95\%$ confidence upper limit of $< 1.7\%$ on planets with periods between 1 and 5 days and radii distributed uniformly between $0.78$ and $0.97 R_J$.  Similarly \citet{Fressin.07} found $f_{VHJ} = 0.18\%$ and $f_{HJ} = 0.29\%$ for the OGLE survey, and they found that the fraction of stars bearing planets with periods less than 2 days is $0.079^{+0.066}_{-0.040}\%$ with $90\%$ confidence error bars. Note that the frequency of HJ is lower for the OGLE survey than the RV surveys because the RV surveys tend to be biased towards higher metallicity stars whereas the OGLE survey is not. Our upper limits on the VHJ and HJ occurence frequencies for field stars are consistent with the frequencies determined by both of these groups.

We can also compare our results to the results from the SWEEPS survey. \citet{Sahu.06} find that $\sim 0.4^{+0.4}_{-0.2}\%$ of stars larger than $0.44~M_{\odot}$ host a Jupiter-sized planet with $P < 4.2~{\rm days}$, assuming all 16 of their candidates are real, which is consistent with the limits that we set. Focusing specifically on the EHJ period range, we note that \citet{Sahu.06} found 5 candidate planets in this range and estimate that at least $\sim 2$ of them are likely to be real. Assuming all 16 candidates are real planets, and that 5 of them are EHJ, their results suggest that $\sim 0.13\%$ of stars host an EHJ. If, on the other hand, we estimate that only half of the candidates are real, and that 2 of them are EHJ, the frequency of EHJ would be $\sim 0.05\%$. Note that we have assumed that the detection probability is constant over the entire period range. Accounting for the enhanced probability of detecting EHJ over VHJ and HJ may lower the frequency estimates of these planets by as much as a factor of $\sim 5$. The SWEEPS frequency of EHJ is below the $95\%$ upper limit that we set assuming 70127 is not a real candidate. If 70127 is real, our EHJ frequency of $0.09^{+0.2}_{-0.077}\%$ (with $1\sigma$ errorbars) would be consistent with the SWEEPS results.

\subsection{Expected Planet Yield}

Based on the planet distribution determined by other surveys we can estimate the expected yield of our survey. \citet{Cumming.08} find that the distribution of planet masses down to $\sim 0.1 M_{J}$ is given by
\begin{equation}
\frac{d^{2}N}{d\ln M d\ln P} = C M^{\alpha} P^{\beta}
\end{equation}
where $\alpha = -0.31 \pm 0.2$ and $\beta = 0.26 \pm 0.1$, $M$ is in Jupiter masses, $P$ is in days, and $C$ is a constant. For notational simplicity, in the following discussion we do not write factors of $M_{J}$ and $R_{J}$. We will adopt the mass dependence from this relation, but not the period dependence, since this relation does not account for the pile-up of planets at short periods. For the period dependence we adopt a constant $C_{\bar{P}}$ that is appropriate for each period range $\bar{P}$ (i.e. EHJ, VHJ or HJ). The model distribution is then given by
\begin{equation}
\frac{dN_{\bar{P}}}{d\ln M} = C_{\bar{P}} M^{\alpha}.
\end{equation}
where
\begin{equation}
C_{\bar{P}} = \frac{f_{\bar{P}}\alpha}{M_{max}^{\alpha} - M_{min}^{\alpha}},
\end{equation}
and $f_{\bar{P}}$ is the fraction of stars with planets in period range $\bar{P}$ with masses between $M_{min} = 0.1$ and $M_{max} = 10$.

\begin{figure}[p]
\plotone{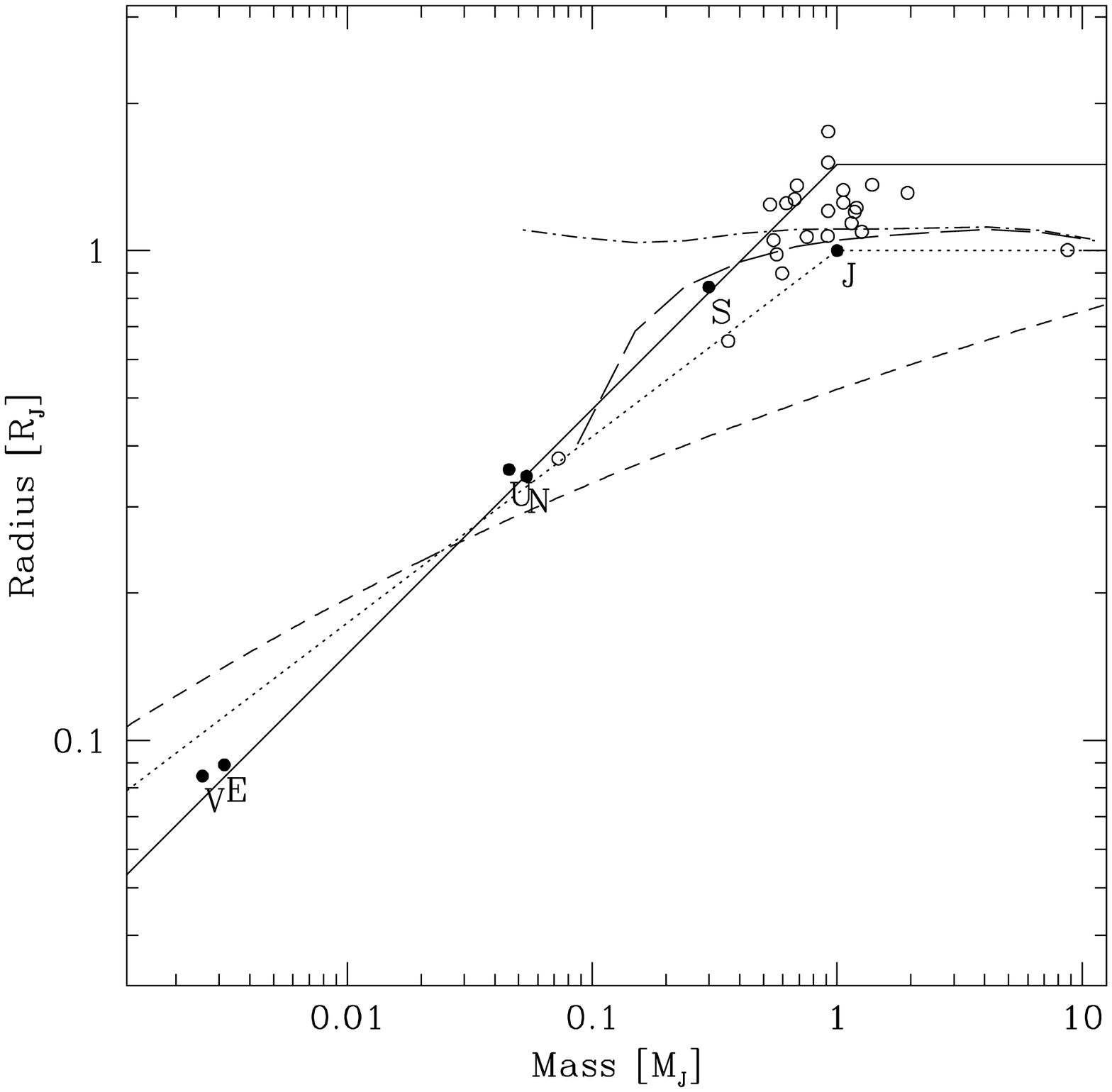}
\caption{The masses and radii of solar system objects (filled circles) and transiting planets (open circles) are compared with two simple mass-radius relations that we adopt for estimating the expected planet yield. The solid line is a power-law of the form $R/R_{J} = 1.5(M/M_{J})^{0.5}$ for $M < M_{J}$ and $R = 1.5R_{J}$ for $M \geq M_{J}$, the dotted line is a power-law of the form $R/R_{J} = 1.0(M/M_{J})^{0.38}$ for $M < M_{J}$ and $R = R_{J}$ for $M \geq M_{J}$. The transiting planet data was taken from \citet{Torres.08}. For comparison we also show theoretical mass-radius relations for a pure ice planet (dashed line), a gas giant with a $25M_{\oplus}$ core (long dashed line), and a pure gas giant (dot-dashed line) from \citet{Fortney.07}. The gas giant planet relations are for a planetary semi-major axis of $0.045$ AU.}
\label{fig:massradius}
\end{figure}

We assume a simple planetary mass-radius relation of the form
\begin{equation}
R(M) = \left\{ \begin{array}{ll}
R_{0}M^{\gamma}, & M < 1.0 \\
R_{0}, & M \geq 1.0
\end{array}
\right.
\end{equation}
with $\gamma = 0.5$, $R_{0} = 1.5$, or $\gamma = 0.38$, $R_{0} = 1$ (see figure~\ref{fig:massradius}). We consider two different mass-radius relations to determine the sensitivity of our results to the assumed relation. The expected number of planet detections in the EHJ, VHJ and HJ period ranges is then given by:
\begin{equation}
N_{\bar{P}} = \sum_{i} \left( \int_{R_{min}}^{R_0} \gamma^{-1} C_{\bar{P}} \left( \frac{R}{R_{0}} \right) ^{\alpha/\gamma} R^{-1} P_{det,\bar{P},i}(R) dR + P_{det,\bar{P},i}(R_{0})C_{\bar{P}}\alpha^{-1}(M_{max}^{\alpha} - 1) \right)
\end{equation}
where the sum is over all stars in the sample, $P_{det,\bar{P},i}(R)$ is the planet detection probability for star $i$ for planets with radii $R$ restricted to the period range $\bar{P}$, and $R_{min}$ is a minimum planetary radius below which $P_{det} = 0$ for all stars (we take $R_{min} = 0.3$). Note that we assume that the planetary mass and period distribution is independent of stellar mass. 

We expect that the sample of stars surveyed by the RV planet searches, which is biased towards high metallicity, provides a better match to the metallicity of the cluster than the stars surveyed by the OGLE transit search, whereas the OGLE sample provides a better match to the metallicity of the field stars than the RV sample. We therefore use the frequencies of HJ and VHJ planets from \citet{Cumming.08} for the cluster and the frequencies from \citet{Gould.06} for the field. We take the frequency of EHJ inferred from the SWEEPS survey for both the field and the cluster. Fixing $C_{HJ}$, $C_{VHJ}$ and $C_{EHJ}$ for cluster members such that the fraction of stars with HJ, VHJ and EHJ larger than $0.1 M_{\odot}$ is $0.012$, $0.0017$, and $0.0005$ respectively, we find $C_{HJ} = 2.4 \times 10^{-3}$, $C_{VHJ} = 3.4 \times 10^{-4}$ and $C_{EHJ} = 1.0 \times 10^{-4}$. For the field stars we use the frequencies from \citet{Gould.06} to set $C_{HJ} = 6.2 \times 10^{-4}$ and $C_{VHJ} = 2.8 \times 10^{-4}$. 

Using the above model and $P_{det}(R)$ values from selection criteria set 2, we find that for cluster members we would expect to detect $\sim 0.10-0.12$ EHJ planets, $0.11-0.16$ VHJ planets, and $0.23-0.35$ HJ planets where the range depends on the assumed mass-radius relation. For field stars, on the other hand, the expected number of detections are $\sim 0.34 - 0.51$, $0.30 - 0.54$ and $0.18 - 0.39$. We conclude that for the above model we would have expected to detect $\sim 1.3 - 2$ stars in our entire survey, and therefore our observations are consistent with the model. Finally, we note that \citet{Beatty.08} predict that there are $\sim 6$ transiting HJ and VHJ planets per square degree orbiting Sun-like stars with $V \leq 20$ at Galactic latitude $b = 3.1^{\circ}$ (see figure 8 of that paper). For our $0.16$ square-degree survey they would predict $\sim 1$ transiting planet detection, which is comparable to the expected planet yield around field stars from our simulations.

\section{Discussion}\label{sec:discussion}

We have presented the results of a deep $\sim 20$ night survey for transiting hot planets in the open cluster M37. This survey stands out from previous ground-based transit surveys both in terms of the size of the telescope used, and the photometric precision attained. We observed $\sim 1450$ cluster members with masses between $0.3~M_{\odot} \la M \la 1.3~M_{\odot}$ as well as 7814 Galactic field stars with masses between $0.1~M_{\odot} \la M \la 2.1~M_{\odot}$. While no candidate planets were found among the cluster members, we did identify one candidate extremely hot Jupiter with a period of $0.77~{\rm days}$ transiting a Galactic field star. However, the follow-up spectroscopic observations needed to confirm the planetary nature of this candidate would be difficult (although perhaps not impossible) to obtain with current technology, given the faintness of the source. We note that if this candidate is real, then we conclude that $0.09^{+0.2}_{-0.077}\%$ of FGKM stars have Jupiter-sized planets with periods between 0.5 and 1.0 days. This result would be consistent with the results from the SWEEPS survey, and would confirm this new class of ultra short period planets. We also note that this planet frequency is small enough that these planets could have escaped detection in most other radial velocity and transit surveys.

The primary result of this survey is an upper limit on the frequency of planets smaller than $1.0R_{J}$. For cluster members, we find that at $95\%$ confidence $< 25\%$ of stars have planets with radii as small as $0.35 R_{J}$ and periods shorter than $1.0~{\rm day}$, $< 44\%$ of stars have planets with radii as small as $0.45 R_{J}$ and periods between $1.0$ and $3.0~{\rm days}$, and $< 49\%$ of stars have planets with radii as small as $0.6 R_{J}$ and periods between $3.0$ and $5.0~{\rm days}$. The upper limits on the smallest planets may be as much as a factor of $50\%$ higher if all the variable stars near the cluster main sequence are cluster members. For the field stars we are able to place $95\%$ confidence upper limits of $16\%$ on the fraction of stars with planets as small as $0.3 R_{J}$ with periods less than $1.0~{\rm day}$, $8.8\%$ on the fraction of stars with planets as small as $0.5 R_{J}$ with periods between $1.0$ and $3.0~{\rm days}$ and $47\%$ on the fraction of stars with planets as small as $0.5 R_{J}$ with periods between $3.0$ and $5.0~{\rm days}$. We estimate that these upper limits may be higher by at most a factor of $\sim 11\%$ due to binarity. While these limits do not approach the observed frequency of Jupiter-sized planets with similar periods, they do represent the first limits on the frequency of planets as small as Neptune. We can now state empirically that extremely hot Neptunes (periods shorter than $1~{\rm day}$) are not ubiquitous, nor are very hot planets with radii intermediate between Neptune and Saturn.

The limits that we place on Jupiter-sized planets are more stringent than previous open cluster transit surveys, but are still above the frequencies measured by RV and Galactic field transit surveys. The primary limitation on open cluster transit surveys appears to be the paucity of stars in these systems. To place a limit on the frequency of HJ that is less than $2\%$ with the same set of observations, M37 would have to have been $\sim 4$ times richer than it is. We also note that for a relatively young cluster like M37, variability may reduce the detectability of Neptune-sized planets by as much as $\sim 50\%$. 

\acknowledgements

We are grateful to C.~Alcock for providing partial support for this
project through his NSF grant (AST-0501681). Funding for M.~Holman
came from NASA Origins grant NNG06GH69G. We would like to thank
G.~F\"{u}r\'{e}sz and A.~Szentgyorgyi for help in preparing the
Hectochelle observations, C.~Burke for a helpful discussion, and the
staff of the MMT, without whom this work would not have been
possible. We would also like to thank the anonymous referee for
several suggestions which improved the quality of this paper, and the
MMT TAC for awarding us a significant amount of telescope time for
this project. This research has made use of the WEBDA database,
operated at the Institute for Astronomy of the University of Vienna;
it has also made use of the SIMBAD database, operated at CDS,
Strasbourg, France.

\clearpage

\begin{deluxetable}{rrrr}
\tabletypesize{\scriptsize}
\tablewidth{0pc}
\tablecaption{Parameters from fitting the noise model in equation~\ref{eqn:noisemodel} to the median RMS-$r$ relations after binning the light curves on several time-scales.}
\tablehead{
\colhead{Binning Time-Scale (minutes)} &
\colhead{$z$ (mag)} &
\colhead{$\sigma_{r}$ (mmag)} &
\colhead{$m_{\sigma_{r}}$ (mag)\tablenotemark{a}}
}
\startdata
   5.0 & 32.45 & 1.35 & 17.57 \\
  10.0 & 32.81 & 1.21 & 17.68 \\
  30.0 & 33.46 & 1.04 & 17.90 \\
  60.0 & 33.95 & 1.00 & 18.21 \\
 120.0 & 34.44 & 0.92 & 18.43 \\
 180.0 & 34.64 & 0.90 & 18.53 \\
 240.0 & 34.63 & 0.93 & 18.58 \\
1440.0 & 35.28 & 0.85 & 18.88 \\
2880.0 & 37.38 & 0.47 & 19.37 \\
7200.0 & 37.91 & 0.33 & 19.23 \\
\cutinhead{$m_{s} = 18.63~{\rm mag}$}
\enddata
\tablenotetext{a}{The $r$ magnitude of a source which has equal red noise and white noise when the light curve is binned at the specified time-scale.}
\label{tab:RMSmodel}
\end{deluxetable}

\begin{deluxetable}{lrrrrrrrrccrrrrr}
\tabletypesize{\scriptsize}
\rotate
\tablewidth{0pc}
\tablecaption{Candidate Transiting Planets}
\tablehead{
\colhead{ID} & 
\colhead{RA} &
\colhead{DEC} &
\colhead{$g$\tablenotemark{a}} &
\colhead{$r$\tablenotemark{a}} &
\colhead{$i$\tablenotemark{a}} &
\colhead{$B$\tablenotemark{b}} &
\colhead{$V$\tablenotemark{b}} &
\colhead{$I$\tablenotemark{c}} &
\colhead{Selection\tablenotemark{d}} &
\colhead{LC\tablenotemark{e}} &
\colhead{Period} &
\colhead{MJD$_{0}$} &
\colhead{Depth} &
\colhead{$q$} &
\colhead{S/N\tablenotemark{f}} \\
& \multicolumn{1}{c}{(J2000)} & \multicolumn{1}{c}{(J2000)} & & & & & & & & & \multicolumn{1}{c}{(days)} & & \multicolumn{1}{c}{(mag)} & & }
\startdata
 30137 & 05:52:53.99 & +32:39:11.7 & 20.86 & 19.74 & 18.95 & 21.20 & 19.94 & 18.50 & 011000111 & 2 &  0.595950 & 53725.32142 & 0.0423 & 0.085 & 24.24 \\
 60161 & 05:53:18.26 & +32:29:46.7 & 20.42 & 19.57 & 19.11 & 20.97 & 19.91 & 18.67 & 001000000 & 2 &  2.623233 & 53726.80688 & 0.0484 & 0.045 & 11.79 \\
 70127 & 05:53:18.46 & +32:28:33.3 & 20.47 & 19.58 & 19.02 & 21.00 & 19.85 & 18.57 & 010000110 & 1 &  0.773530 & 53725.31696 & 0.0197 & 0.060 & 10.57 \\
 80009 & 05:53:02.04 & +32:23:27.6 & 15.75 & 15.09 & 14.76 & 16.13 & 15.30 & 14.32 & 000000001 & 3 &  1.519846 & 53724.56637 & 0.0483 & 0.045 & 12.29 \\
 80014 & 05:53:12.12 & +32:23:51.2 & 16.28 & 15.77 & 15.46 & 16.62 & 15.92 & 15.03 & 001001000 & 2 &  0.588375 & 53725.61226 & 0.0054 & 0.100 & 13.22 \\
 90279 & 05:52:53.06 & +32:22:57.0 & 23.58 & 21.87 & 20.45 & 24.37 & 22.43 & 19.96 & 111000111 & 2 &  1.141717 & 53726.05590 & 0.3173 & 0.045 & 37.23 \\
110021 & 05:52:26.24 & +32:40:44.6 & 16.36 & 15.77 & 15.43 & 16.77 & 15.99 & 14.99 & 000010000 & 3 &  3.535574 & 53727.68767 & 0.1215 & 0.035 & 13.35 \\
120050 & 05:52:46.14 & +32:39:33.6 & 17.71 & 16.98 & 16.55 & 18.20 & 17.35 & 16.11 & 111111000 & 3 &  1.847414 & 53725.76532 & 0.0346 & 0.045 & 14.90 \\
160017 & 05:52:23.89 & +32:28:26.3 & 16.31 & 15.60 & 15.24 & 16.75 & 15.90 & 14.81 & 000110000 & 3 &  1.560626 & 53724.96524 & 0.0141 & 0.055 &  4.37 \\
160311 & 05:52:19.64 & +32:26:50.4 & 23.20 & 21.63 & 20.64 & 23.78 & 22.26 & 20.17 & 111000111 & 2 &  2.027186 & 53726.18244 & 0.1789 & 0.025 & 26.33 \\
170049 & 05:52:44.65 & +32:25:55.3 & 18.36 & 17.65 & 17.30 & 18.74 & 17.90 & 16.87 & 000000001 & 2 &  2.261081 & 53725.22479 & 0.0181 & 0.055 & 10.29 \\
170100 & 05:52:38.44 & +32:23:29.5 & 20.34 & 18.73 & 17.80 & 21.02 & 19.43 & 17.34 & 000001000 & 3 &  2.746378 & 53725.19817 & 0.2963 & 0.025 & 20.78 \\
230082 & 05:52:01.32 & +32:32:51.0 & 17.71 & 17.08 & 16.71 & 18.15 & 17.36 & 16.27 & 011000011 & 2 &  3.335057 & 53726.43670 & 0.0280 & 0.025 & 15.05 \\
270149 & 05:51:59.04 & +32:20:38.0 & 20.59 & 19.47 & 18.87 & 21.68 & 19.98 & 18.42 & 000000111 & 3 &  1.097491 & 53724.89160 & 0.2132 & 0.065 & 11.11 \\
280287 & 05:51:37.41 & +32:45:03.0 & 23.80 & 22.12 & 20.78 & 24.68 & 22.82 & 20.29 & 111000011 & 2 &  0.456034 & 53725.41220 & 0.1331 & 0.055 & 17.35 \\
330224 & 05:51:30.37 & +32:30:29.3 & 21.64 & 20.46 & 19.91 & 22.11 & 20.94 & 19.46 & 010000110 & 1 &  0.648749 & 53725.20636 & 0.0502 & 0.060 & 13.79 \\
\enddata
\tablenotetext{a}{The magnitude of the source in the photometric catalog (Paper I)}
\tablenotetext{b}{Value from \citet{Kalirai.01}}
\tablenotetext{c}{Value from transforming $ri$ photometry to the $I_{C}$ photometry of \citet{Nilakshi.02}}
\tablenotetext{d}{9-bit flag to indicate which selection criteria sets and resolutions selected the candidate. From the left, the first three bits indicate if the candidate was selected by criteria sets 1, 2 and 3 respectively, using a high resolution BLS search. The middle three bits are for the low resolution BLS search applied to candidate cluster members, while the last three bits are for the low resolution BLS search applied to field stars.}
\tablenotetext{e}{This integer indicates if the BLS parameters in this table were determined for: 1. the fully processed, trend-filtered light curve, 2. the light curve processed through steps 1 and 2 of \S~\ref{sec:lcproc} only, or 3. the un-processed light curve.}
\tablenotetext{f}{Signal to Pink noise ratio (eq.~\ref{eqn:sigtopink}).}
\label{tab:transitcandidates}
\end{deluxetable}

\clearpage

\begin{deluxetable}{lccrrrrrrrrrrrr}
\tabletypesize{\scriptsize}
\rotate
\tablewidth{0pc}
\tablecaption{$95\%$ Upper limits on the Planet Occurrence Frequency}
\tablehead{
 &
\colhead{Period} &
\colhead{Selection} &
\colhead{$0.3R_{J}$} &
\colhead{$0.35R_{J}$} &
\colhead{$0.4R_{J}$} &
\colhead{$0.45R_{J}$} &
\colhead{$0.5R_{J}$} &
\colhead{$0.6R_{J}$} &
\colhead{$0.7R_{J}$} &
\colhead{$0.8R_{J}$} &
\colhead{$0.9R_{J}$} &
\colhead{$1.0R_{J}$} &
\colhead{$1.5R_{J}$} &
\colhead{$2.0R_{J}$}
}
\startdata
Cluster & EHJ & 1 & 1.000 & 0.421 & 0.139 & 0.067 & 0.042 & 0.023 & 0.017 & 0.015 & 0.014 & 0.014 & 0.026 & 0.048 \\
& & 2 & 1.000 & 0.247 & 0.091 & 0.049 & 0.032 & 0.019 & 0.015 & 0.012 & 0.011 & 0.011 & 0.012 & 0.025 \\
& & 3 & $\cdots$ & $\cdots$ & $\cdots$ & $\cdots$ & $\cdots$ & $\cdots$ & $\cdots$ & $\cdots$ & $\cdots$ & 0.014 & 0.010 & 0.014 \\
& VHJ & 1 & 1.000 & 1.000 & 1.000 & 0.754 & 0.328 & 0.117 & 0.067 & 0.048 & 0.040 & 0.036 & 0.050 & 0.100 \\
& & 2 & 1.000 & 1.000 & 1.000 & 0.444 & 0.213 & 0.086 & 0.051 & 0.037 & 0.031 & 0.027 & 0.023 & 0.056 \\
& & 3 & $\cdots$ & $\cdots$ & $\cdots$ & $\cdots$ & $\cdots$ & $\cdots$ & $\cdots$ & $\cdots$ & $\cdots$ & 0.044 & 0.026 & 0.031 \\
& HJ & 1 & 1.000 & 1.000 & 1.000 & 1.000 & 1.000 & 0.967 & 0.416 & 0.248 & 0.186 & 0.153 & 0.165 & 0.300 \\
& & 2 & 1.000 & 1.000 & 1.000 & 1.000 & 1.000 & 0.485 & 0.226 & 0.139 & 0.104 & 0.083 & 0.069 & 0.149 \\
& & 3 & $\cdots$ & $\cdots$ & $\cdots$ & $\cdots$ & $\cdots$ & $\cdots$ & $\cdots$ & $\cdots$ & $\cdots$ & 0.202 & 0.104 & 0.124 \\
Field & EHJ & 1 & 0.230 & $\cdots$ &$\cdots$ &$\cdots$ &0.023 & $\cdots$ &0.007 & $\cdots$ &$\cdots$ &0.004 & 0.003 & 0.004 \\
& & 2 & 0.158 & $\cdots$ &$\cdots$ &$\cdots$ &0.014 & $\cdots$ &0.005 & $\cdots$ &$\cdots$ &0.003 & 0.002 & 0.002 \\
& & 3 & $\cdots$ & $\cdots$ & $\cdots$ & $\cdots$ & $\cdots$ & $\cdots$ & $\cdots$ & $\cdots$ & $\cdots$ & 0.003 & 0.002 & 0.002 \\
& VHJ & 1 & 1.000 & $\cdots$ &$\cdots$ &$\cdots$ &0.148 & $\cdots$ &0.035 & $\cdots$ &$\cdots$ &0.012 & 0.007 & 0.007 \\
& & 2 & 1.000 & $\cdots$ &$\cdots$ &$\cdots$ &0.088 & $\cdots$ &0.020 & $\cdots$ &$\cdots$ &0.008 & 0.005 & 0.004 \\
& & 3 & $\cdots$ & $\cdots$ & $\cdots$ & $\cdots$ & $\cdots$ & $\cdots$ & $\cdots$ & $\cdots$ & $\cdots$ & 0.009 & 0.005 & 0.005 \\
& HJ & 1 & 1.000 & $\cdots$ &$\cdots$ &$\cdots$ &1.000 & $\cdots$ &0.240 & $\cdots$ &$\cdots$ &0.062 & 0.028 & 0.026 \\
& & 2 & 1.000 & $\cdots$ &$\cdots$ &$\cdots$ &0.469 & $\cdots$ &0.095 & $\cdots$ &$\cdots$ &0.027 & 0.013 & 0.012 \\
& & 3 & $\cdots$ & $\cdots$ & $\cdots$ & $\cdots$ & $\cdots$ & $\cdots$ & $\cdots$ & $\cdots$ & $\cdots$ & 0.044 & 0.020 & 0.019 \\
\enddata
\label{tab:f95vsRadius}
\end{deluxetable}

\begin{deluxetable}{lrrrr}
\tabletypesize{\scriptsize}
\tablewidth{0pc}
\tablecaption{$95\%$ Upper limits on the Planet Occurrence Frequency for Field Stars Using Different Galactic Models}
\tablehead{
&
\colhead{Radius [$R_{J}$]} &
\colhead{EHJ\tablenotemark{a}} &
\colhead{VHJ} &
\colhead{HJ}
}
\startdata
Trilegal Model & 0.3 & 0.143 & $\cdots$ & $\cdots$ \\
               & 0.5 & 0.014 & 0.082 & 0.401 \\
               & 0.7 & 0.005 & 0.022 & 0.104 \\
               & 1.0 & 0.003 & 0.009 & 0.033 \\
               & 1.5 & 0.002 & 0.005 & 0.015 \\
               & 2.0 & 0.002 & 0.005 & 0.013 \\
Besan\c{c}on Model & 0.3 & 0.151 & $\cdots$ & $\cdots$ \\ 
$A_{V} = 0.5~{\rm mag}/{\rm kpc}$ & 0.5 & 0.010 & 0.065 & 0.394 \\
               & 0.7 & 0.004 & 0.016 & 0.070 \\
               & 1.0 & 0.003 & 0.007 & 0.023 \\
               & 1.5 & 0.002 & 0.005 & 0.014 \\
               & 2.0 & 0.002 & 0.005 & 0.015 \\
\enddata
\tablenotetext{a}{The $95\%$ upper limit on the occurence frequency of planets with the specified Radius and within the EHJ period range.}
\label{tab:f95galmodel}
\end{deluxetable}

\begin{deluxetable}{rrrr}
\tabletypesize{\scriptsize}
\tablewidth{0pc}
\tablecaption{$95\%$ Upper limits on the Planet Occurrence Frequency for Cluster Stars Assuming All Variables are Cluster Members}
\tablehead{
\colhead{Radius [$R_{J}$]} &
\colhead{EHJ\tablenotemark{a}} &
\colhead{VHJ} &
\colhead{HJ}
}
\startdata
0.35 & 0.381 & 1.000 & 1.000 \\
0.40 & 0.123 & 1.000 & 1.000 \\
0.45 & 0.060 & 0.633 & 1.000 \\
0.50 & 0.037 & 0.281 & 1.000 \\
0.60 & 0.021 & 0.103 & 0.636 \\
0.70 & 0.015 & 0.058 & 0.277 \\
0.80 & 0.013 & 0.040 & 0.161 \\
1.00 & 0.011 & 0.028 & 0.090 \\
1.50 & 0.012 & 0.024 & 0.071 \\
2.00 & 0.022 & 0.056 & 0.154 \\
\enddata
\tablenotetext{a}{The $95\%$ upper limit on the occurence frequency of planets with the specified Radius and within the EHJ period range.}
\label{tab:f95vars}
\end{deluxetable}

\end{document}